\documentclass[letterpaper, 10pt, conference]{ieeeconf}      % Use this line for a4
\IEEEoverridecommandlockouts                              % This command is only
\usepackage{graphicx}
\usepackage{amsmath}
\usepackage{amssymb}
\usepackage{float}

\title{\LARGE \bf
Multi Time Scale Behaviour of The Formation of Multiple Groups of Nonholonomic Wheeled Mobile Robots 
}

\author{Soumic Sarkar$^{1}$ and Indra Narayan Kar$^{2}$% <-this % stops a space
\thanks{$^{1}$S. Sarkar is with Department of Electrical Engineering, Indian Institute of Technology Delhi, New Delhi-110 016, INDIA
        {\tt\small soumic4it at gmail.com}}%
\thanks{$^{2}$I.N. Kar is with the Faculty of Electrical Engineering, Indian Institute of Technology Delhi, New Delhi-110 016, INDIA
        {\tt\small ink at iit.ac.in}}%
}

\begin{document}

\maketitle
\thispagestyle{empty}
\pagestyle{empty}

%%%%%%%%%%%%%%%%%%%%%%%%%%%%%%%%%%%%%%%%%%%%%%%%%%%%%%%%%%%%%%%%%%%%%%%%%%%%%%%%
\begin{abstract}
Different geometric patterns and shapes are generated using groups of agents, and this needs formation control. In this paper, Centroid Based Transformation (CBT), has been applied to decompose the combined dynamics of nonholonomic Wheeled Mobile Robots (WMRs) into three subsystems: intra and inter group shape dynamics, and the dynamics of the centroid. The intra group shape dynamics can further be partitioned into the shape dynamics of each group, giving the notion of multiple group. Thus separate controllers have been designed for each subsystem. The gains of the controllers are such chosen that the overall system becomes singularly perturbed system, and different subsystems converge to their desired values at different times. Then multi-time scale convergence analysis has been carried out in this paper. Negative gradient of a potential based function has been added to the controller to ensure collision avoidance among the robots. Simulation results have been provided to demonstrate the effectiveness of the proposed controller. 
\end{abstract}

%%%%%%%%%%%%%%%%%%%%%%%%%%%%%%%%%%%%%%%%%%%%%%%%%%%%%%%%%%%%%%%%%%%%%%%%%%%%%%%%
\section{INTRODUCTION}

The study of the collective behavior of birds, animals, fishes, etc. has not only drawn the attention of biologists, but also of computer scientists and roboticists. Thus several methods of cooperative control \cite{c10} of multi-agent system have evolved, where a single robot is not sufficient to accomplish the given task, like navigation and foraging of unknown territory. These methods can broadly be categorized as the behavior based approach (\cite{c1}-\cite{c3}), leader follower based approach \cite{c4}-\cite{c5}, virtual structure based approach \cite{c6}-\cite{c9}, artificial potential based navigation \cite{c13}-\cite{c15}, graph theoretic method \cite{c11}-\cite{c12}, formation shape control \cite{c16}-\cite{c21}. Among other works carried out on single group of robots, cluster space control \cite{c34}, distance based formation \cite{c35}, formation control of nonholonomic robots \cite{c4}, kinematic control \cite{c27}, and mobile robots subject to wheel slip \cite{c36}, segregation of heterogeneous robots \cite{c37}, are to name a few.\\
The problem associated with the formation control of multi-agent system is that it becomes difficult to accurately position the robot within the group, as the number of robots increase. To address this issue shape control and region based shape control have been proposed, such that the robots form a desired shape during movement. The desired shape can be union or intersection of different geometric shapes. Region based shape control has been extended to multiple groups of robots \cite{c22}-\cite{c24}. However, the robots can stay anywhere inside the specified region without colliding with each other. This means that the position of a robot inside a group can be specified and can further be controlled. Therefore the position of a group of robots inside a larger group of robots can also be specified and controlled. Moreover, when it comes to the control of multiple groups of robots, there should be at least one robot to convey the information of that group to another group.\\
In an attempt to solve the aforementioned problem, a hierarchical multi level topology have been proposed, here in this paper, which is based on the centroid based transformations \cite{c16}-\cite{c19} for single group of robots. In this architecture, the large group of robots has been partitioned into relatively small basic units containing three robots. Then the centroids of each unit are connected to form larger module containing more robots. Extending the process gives a hierarchical architecture which is a composition of relatively smaller modules. As the construction of this topology involves connecting the centroid, it is named Centroid Based Topology (CBT). The CBTs basically capture the constraint relationship among the robots. Another advantage of CBT is that it separates shape variables from the centroid and thus separates the formation shape controller and tracking controller design. As the centroid moves, the entire structure moves maintaining the shape specified by the shape variables. In \cite{c39}, different CBTs for multiple groups of robots have been introduced, to get a modular architecture distinguishable in the form of \textbf{intra group shape variables}, \textbf{inter group shape variables} along with centroid. It is to be noted that a centroid based decoupling approach is adopted in \cite{c40} to improve the convergence rate of cyclic pursuit scheme for vehicle networks. However, the modularity presented in this paper allows one to distinguish even the intra group shape variables of different groups. Thus multi time scale convergence behaviour of singular perturbation approach can be utilized in this modular framework so that convergence times of different groups can be selected based on the choice of the user. Based on this modular structure, a novel feedback control algorithm is proposed here in this paper. The gains of the feedback controller are so selected that the closed loop dynamics becomes singularly perturbed system. Thus the model of the system reduces to the dynamics of centroid after the convergence of intra and inter group shape dynamics. This technique allows us to give importance to the part of the formation dynamics, which has to converge earlier than the others, based on the choice of the gains of the respective controller. Potential function based controller has also been designed to avoid inter robot collision.

\section{PROBLEM FORMULATION}
Given a set of $N$ robots with nonholonomic constraint \cite{c25} - \cite{c26}, given by
\begin{equation}
\label{801}
\ddot{p}_i=A_i(\theta_i,\dot{\theta}_i)\dot{p}_i+B_i(\theta_i)u_i+C_i(\dot{\theta}_i)
\end{equation}
where
$$
A_i(\theta_i,\dot{\theta}_i)=\begin{bmatrix}
-\sin\theta_i \cos\theta_i \dot{\theta}_i & -\sin^2\theta_i \dot{\theta}_i \\
\cos^2\theta_i \dot{\theta}_i & \sin\theta_i \cos\theta_i \dot{\theta}_i
\end{bmatrix}
$$
\[
B_i(\theta_i)=\begin{bmatrix}
\frac{\cos\theta_i}{m_1r}-\frac{dR\sin\theta_i}{Jr} & \frac{\cos\theta_i}{m_1r}+\frac{dR\sin\theta_i}{Jr} \\
\frac{\sin\theta_i}{m_1r}+\frac{dR\cos\theta_i}{Jr} & \frac{\sin\theta_i}{m_1r}-\frac{dR\cos\theta_i}{Jr}
\end{bmatrix}
\]
$$
C_i(\dot{\theta}_i)=\begin{bmatrix}
-d\dot{\theta}_i^2\cos\theta_i \\
-d\dot{\theta}_i^2\sin\theta_i
\end{bmatrix}
$$
$$
u_i=[\tau_{ri},\tau_{li}]^T
$$
And 
$$
J\ddot{\theta}=\frac{R}{r}(\tau_r-\tau_l)
$$
where, $m_1$ is the mass of robot, $J=I-m_1d^2$, $I$ is moment of inertia, $R$ is the distance between left and right wheels, $r$ is the radius of each wheel, $d$ is the distance from wheel axis to the center of mass, and $\theta$ is the orientation. The positions of the robots are described by $p_i=[x_i,y_i]^T$, $i=1,2,\ldots,N$ in the inertial coordinate frame, and $u_i=[\tau_r,\tau_l]^T$ is the control torque input. Then, for a single group of robots, a linear transformation $\Phi\in\mathbb{R}^{2N\times2N}$, can be defined, that produces the following matrix equation 
\begin{equation}
\label{eq1}
[z_1^T,z_2^T,\ldots,z_{N-1}^T,z_c^T]^T=\Phi[p_1^T,p_2^T,\ldots,p_N^T]^T
\end{equation}
where $z_i=[z_{xi},z_{yi}]^T\in\mathbb{R}^{2\times 1}$, $i=1,2,\ldots,(N-1)$ are the shape defining vectors or shape variables in transformed coordinate, and these vectors define the geometric shape of formation of swarms. 
Clearly, the transformation $\Phi$ generates shape variables along with the centroid for a single group of robots. \\
For multiple groups of robots, these shape variables can be categorized into two parts. The shape variables which represent the shape of each subgroup, are \textbf{intra group shape variables}. However, the variables which describe the interconnection among the groups, each group being considered as a single agent, concentrated onto the centroid of that group,  are \textbf{inter group shape variables}.
Suppose there are $m$ subgroups and each subgroup contains $\rho_i$ number of robots, where $i=1,2,...,m$ $(\sum_{i=1}^{m} \rho_i=N$, $N$ being the total number of robots$)$. Then the total number of intra group shape variables is $\rho=\sum_{i=1}^{m} (\rho_i-1)$, and total number of inter group shape variables is $(m-1)$. The intra group shape variables for each subgroup is defined as 
$$Z_j=[z_{j1}^T,z_{j2}^T,\ldots,z_{j(\rho_i-1)}^T]^T$$
where, $Z_j\in \mathbb{R}^{1 \times 2(\rho_i-1)}$, $i,j=1,2,...,m$ and $z_{jk}\in \mathbb{R}^{2 \times 1}$, $k=1,2,...,\rho_i-1$. Therefore, the intra groups shape vectors are defined in a compact form as
$$Z_s=[Z_1^T,Z_2^T,\cdots,Z_m^T]^T$$
The inter group shape variables considering the centroid of each group as an agent, is defined as
$$Z_r=[z_{r1}^T,z_{r2}^T,\ldots,z_{r(m-1)}^T]^T$$
where, $Z_r \in \mathbb{R}^{1 \times 2(m-1)}$ and $z_{ri} \in \mathbb{R}^{2 \times 1}$, $i=1,2,...,(m-1)$.
The geometric center of mass, $z_c$ is, defined by
$$z_c=\frac{1}{N}\sum_{i=1}^{N} p_i $$
Using the above definitions for multiple groups of robots, the intra group, inter group shape variable, and centroid can be written in compact form using a CBT \cite{c39} $\Phi_M$ as
$$[Z_s^T,Z_r^T,z_c^T]^T=\Phi_M[p_1^T,p_2^T,\ldots,p_N^T]^T$$
The detailed description of the matrices $\Phi$ and $\Phi_M$ is given in Section III and IV. 
Define, the desired intra group shape variables $Z_{sd}$, the inter group shape variables $Z_{rd}$, and the desired trajectory of the centroid $z_{cd}$.
%, the desired vectors can be written as
%$$[Z_{sd},Z_{rd},z_{cd}]^T=\Phi_M[p_{1d},p_{2d},\ldots,p_{Nd}]^T$$
%where $p_{id}$ is the desired trajectory of $i^{th}$ robot. \\
Let $Z=[Z_s^T,Z_r^T,z_c^T]^T$ and $X=[p_1^T,p_2^T,\ldots,p_N^T]^T$ and $Z_d=[Z_{sd}^T,Z_{rd}^T,z_{cd}^T]^T$ and $X_d=[p_{1d}^T,p_{2d}^T,\ldots,p_{Nd}^T]^T$. The following equation gives the transformation from Cartesian to the transformed coordinate.
$$Z=\Phi_M X; \ Z_d=\Phi_M X_d$$
Therefore, the convergence of $Z\rightarrow Z_{d}$ as $t \rightarrow \infty$ leads to the convergence of $X\rightarrow X_{d}$ as $t \rightarrow \infty$ as $\Phi_M$ is nonsingular. However, the objective to devise controllers for different dynamics such that they converge to their desired values at different times. Based on this, the formation control problem has been divided into the following sub-problems. \\
\textbf{Intra Group Formation Control}: Given a reference constant $Z_{id}$, determine a control law such that intra group shape variables $Z_{i}(t)$ converges to the desired value as
$$\lim_{t \to t_i} Z_{i}(t)\rightarrow Z_{id}$$
\textbf{Inter Group Formation Control}: Given a reference constant $Z_{rd}$ 
determine a control law such that inter group shape variable $Z_{r}(t)$ converges to the desired value as
$$\lim_{t \to t_r} Z_{r}(t)\rightarrow Z_{rd}$$
\textbf{Trajectory Tracking}: Given a reference time varying trajectory $z_{cd}(t)$ determine a control law such that the centroid $z_c(t)$ converges to the desired trajectory as
$$\lim_{t \to \infty} z_c(t)\rightarrow z_{cd}(t)$$
where, $0<t_i<t_r<t_c<t<\infty$, $i=1,2,...,m$.

\section{GENERAL FORM OF CENTROID BASED TRANSFORMATION}

In centroid based representation \cite{c16}-\cite{c19} of formation of a single group of robots, the centroid is being retained, as it contains all the positional information of the group of robots. All other vectors (\textbf{shape variables}) describe the connectivity relationship among the robots in the group.
However, the general transformation matrix for $N$ robots can be given as
$$
\Phi=\left[\Phi_{r}^T, \Phi_{c}^T\right]^T
$$
where, $I_2$ is the identity matrix of dimension $2$. 
The dimension of the matrix $\Phi_{r}$ is $(2(N-1) \times 2N)$. 
The matrix $\Phi_{c}$ is $(2 \times 2N)$ and it captures the information of the coefficients to generate the centroid vector.
As the centroid is to be retained, the last row of the block matrix $\Phi$ is given by, 
$$
\Phi_{c}=\frac{1}{N}
\begin{bmatrix}
I_2 I_2 \cdots I_2 
\end{bmatrix} \in R^{2 \times 2N} 
$$
\section{TRANSFORMATION MATRIX FOR MULTIPLE GROUPS OF ROBOTS}

This section mainly describes how to generate centroid based transformation matrices for multiple groups of robots. 
\begin{figure}[h]
\centering
\fbox{\includegraphics[width=8cm, height=6cm]{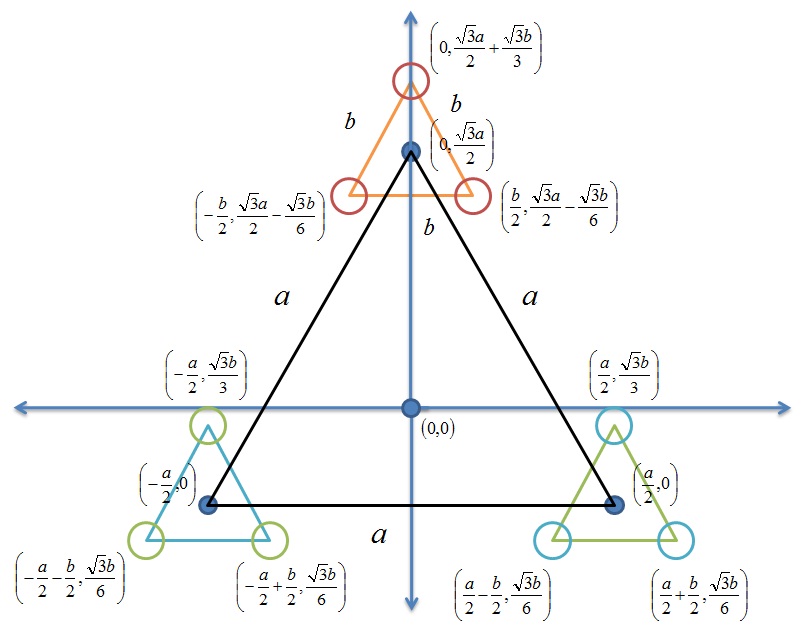}}
\caption{Schematic Representation of Multiple Groups of Robots}
\end{figure}
Fig. 4 depicts nine robots divided in subgroups with three robots in each. The shape variables of the Jacobi transformation applied on the nodes of each subgroup are collectively \textbf{intra group} shape variables. The \textbf{inter group} shape variables can be found applying the Jacobi transformation on the centroids of subgroups. With this modularity even the intra group shape variables of each subgroup is identifiable. The derivation of this example of transformation gives a heuristic understanding and some intuitive feeling for the solution of the stated problem. The intra group shape variables are
$$
\begin{cases}
Z_{1}\Rightarrow\begin{cases} 
z_{11}=\frac{1}{\sqrt{2}}(p_2-p_1) \\
z_{12}=p_3-\frac{1}{2}(p_1+p_2) \\
\end{cases}\\
Z_2\Rightarrow\begin{cases} 
z_{21}= \frac{1}{\sqrt{2}}(p_4-p_5) \\
z_{22}=p_6-\frac{1}{2}(p_4+p_5) \\
\end{cases}\\
Z_3\Rightarrow\begin{cases} 
z_{31}= \frac{1}{\sqrt{2}}(p_7-p_8) \\
z_{32}=p_9-\frac{1}{2}(p_7+p_8) \\
\end{cases}\\
Z_r\Rightarrow\begin{cases}
z_{r1}= \frac{1}{\sqrt{2}}(\mu_1-\mu_2) \\
z_{r2}=\mu_3-\frac{1}{2}(\mu_1+\mu_2) \\
\end{cases}\\
\end{cases}
$$
where, $\mu_1=\frac{1}{3}(p_1+p_2+p_3)$, $\mu_2=\frac{1}{3}(p_4+p_5+p_6)$, $\mu_3=\frac{1}{3}(p_7+p_8+p_9)$.
\subsection{General Form of The Transformation Matrix for Multiple Groups}
%\textbf{General Form of The Transformation Matrix for Multiple Groups}\\
Therefore, the general form of the transformation matrix for multiple groups of robots can be written as follows
$$
\Phi_M=\left[\Phi_1^T,\Phi_2^T,\hdots,\Phi_m^T,\Phi_{r}^T,\Phi_{c}^T\right]^T
$$
Where, $\Phi_1, \Phi_2,...,\Phi_m$ are $(2(\rho_1-1) \times 2N), (2(\rho_2-1) \times 2N),..., (2(\rho_m-1) \times 2N)$ dimensional matrices and $m$ is the total number of subgroups. The scalar $((\rho_i)-1), \ i=1,2,\cdots, m$ are the number of shape variables required to represent $i^{th}$ subgroup. $\Phi_r$ is $(2(m-1) \times 2N)$ and $\Phi_c$ is $(2 \times 2N)$. We write the transformation matrix in a more compact form as
$$
\Phi_M=\left[\mathbf{\Phi_m}^T,\Phi_{r}^T,\Phi_{c}^T\right]^T
$$
Suppose there are $m$ groups of robots and in the $i^{th}$ group, there are $n_i$ number of robots. Then the dimension of the matrix $\mathbf{\Phi_m}$ is $2((\rho_1+\rho_2+\cdots+\rho_m)-m)\times 2N$. Again $\mathbf{\Phi_m}$ can be written in block diagonal form as $\mathbf{\Phi_m}=diag\{\Phi_i\}$, where, $\Phi_i$ denotes the transformation matrix for the $i_{th}$ group of robots containing $\rho_i$ number of robots. The dimension of $\Phi_i$ is $2(\rho_i-1)\times 2\rho_i$.
\subsection{Shape Variable Generation Algorithm for Multiple Groups of Robots}
The following steps describe the generation algorithm for the shape variables of multiple groups of robots.\\
\textbf{step 1}: calculate intra group shape variables for each subgroup, i.e., $Z_1,\cdots, Z_m$. \\
\textbf{step 2}: calculate the centroid of each subgroup, i.e. $\mu_1,\cdots\mu_m$\\
\textbf{step 3}: calculate inter group shape variables for the overall group assuming each group as an agent. \\
\textbf{step 4}: calculate the overall centroid. \\
\textbf{step 5}: write the vectors in matrix form to get the transformation matrix. \\
The algorithm above will generate all the vectors of the transformation. The coefficients of the vectors collectively form the transformation matrix for multiple groups of robots.

\section{FORMATION CONTROLLER DESIGN AND STABILITY ANALYSIS}

\subsection{Formation Dynamics}
The entire formation of $N$ WMRs can be viewed as a deformable body whose shape and movement can be described by vectors in transformed coordinate. 
Define the notation $A_i=A_i(\theta_i,\dot{\theta}_i)$, $B_i=B_i(\theta_i)$ and $C_i=C_i(\dot{\theta}_i)$ for $i=1,2,\ldots,N$, where $\theta_i$ and $\dot{\theta}_i$ are the orientation and angular speed of the $i$-th WMR as described in \eqref{801}.\\ 
The dynamic equation of $N$ WMRs can be written as
\begin{equation}
\label{eq23}
\ddot{X}=\textbf{A}\dot{X}+\textbf{B}U+\textbf{C}
\end{equation}
where, $\textbf{A}=diag\{A_1,A_2,\ldots,A_N\}$, $\textbf{B}=diag\{B_1,B_2,\ldots,B_N\}$, $\textbf{C}=diag\{C_1,C_2,\ldots,C_N\}$ and $U=[u_1^t,u_2^T,...,u_N^T]^T$.
Using the transformation $\Phi_M$, \eqref{eq23} can be written as
\begin{equation}
\label{eq25}
\ddot{Z}=\mathbf{P}\dot{Z}+\mathbf{Q}U+\mathbf{R}
\end{equation}
where $\mathbf{P}=\Phi_M \textbf{A} \Phi_M^{-1}; \ \mathbf{Q}=\Phi_M B U; \ \mathbf{R}=\Phi_M \textbf{C}$. With equation \eqref{eq25}, controllers are designed in the next section such that different dynamics converge to their desired values at different times.

\subsection{Asymptotic Stability Analysis of Three-time Scale Singularly Perturbed Systems}

The asymptotic stability analysis of three-time scale singularly perturbed systems composed of the twice application of two-time scale analysis, given in Appendices I and II. There are two ways to address the analysis: \textit{Top-Down} and \textit{Bottom-Up}. These two approaches logically select the slow and fast dynamics sequentially. The details of these two approaches can be found in \cite{c32}-\cite{c44}. The \textit{Bottom-Up} approach is considered here in this paper to understand the natural evolution of the multi-time scale singularly perturbed systems in their own configuration spaces. A generic three-time scale model can be described by
\begin{equation}
\label{multimod01}
\begin{split}
\dot{x}_1&=f_1(x_1,x_2,x_3), x_1\in R^{m_1} \\
\epsilon_1\dot{x}_2&=f_2(x_1,x_2,x_3), x_2\in R^{m_2} \\
\epsilon_1\epsilon_2\dot{x}_3&=f_3(x_1,x_2,x_3), x_2\in R^{m_3}
\end{split}
\end{equation}
The model \eqref{multimod01} can be sequentially decomposed into two different two-time scale models. The first two-time scale model considers the time scale defined by the stretched time scale $t_{2}=\frac{t}{\epsilon_{1}\epsilon_{2}}$, where the reduced (slow) subsystems are defined by
\begin{equation}
\label{multimod02}
\begin{split}
\dot{x}_1&=f_1(x_1,x_2,F_2(x_1,x_2)) \\
\epsilon_1\dot{x}_2&=f_2(x_1,x_2,F_2(x_1,x_2)) \\
\end{split}
\end{equation}
and the boundary layer subsystem is given by
\begin{equation}
\label{multboundary}
\epsilon_1\epsilon_2\frac{dx_3}{dt}=\frac{dx_3}{dt_{2}}=f_3(x_1,x_2,x_3(t_2))
\end{equation}
where in \eqref{multboundary}, $x_1,x_2$ are treated like fix parameters and $x_3(t_2)$ evolves on its stretched time scale $t_2$. $F_2(x_1,x_2)$ in \eqref{multimod02} represents the quasi-steady-state of the boundary layer \eqref{multboundary}, when $\epsilon_2=0$, that is, $0=f_3(x_1,x_2,x_3)\rightarrow x_3=F_2(x_1,x_2)$. 
From \eqref{multimod02}, the slow system can be written as
\begin{equation}
\label{slowest}
\begin{split}
\dot{x}_1&=f_1(x_1,F_1(x_1),F_2(x_1,F_1(x_1)))
\end{split}
\end{equation}
where $F_1(x_1)$ in \eqref{slowest} is the quasi-steady-state of the boundary layer
\begin{equation}
\label{boundary2}
\epsilon_1\frac{dx_2}{dt}=\frac{dx_2}{dt_1}=f_2(x_1,x_2(t_1),F_2(x_1,x_2(t_1))
\end{equation}
when $\epsilon_1=0$, that is, $0=f_2(x_1,x_2,F(x_1,x_2))\rightarrow x_2=F_1(x_1)$.
A Lyapunov Function is constructed initially to satisfy the growth requirements of \eqref{slowest} and \eqref{boundary2} as in Appendix I and II
\begin{equation}
\label{lyapn-2}
V_{1}(x_1,x_2)=(1-d_{1})V(x_1)+d_{1}W(x_1,x_2)
\end{equation}
where $0<d_{2}<1$, and $V(x_1)$ and $W(x_1,x_2)$ are the chosen Lyapunov function that satisfy the growth requirements for \eqref{slowest} and \eqref{boundary2} respectively. Note that the construction of $V_1(x_1,x_2)$ with the Lyapunov functions for the subsystems $x_1$ and $x_2$ must satisfy the growth requirements on $\overline{F}_2(x_1,x_2,F_2(x_1,x_2))$ and $f_3(x_1,x_2,F_2(x_1,x_2))$. The bound on $\epsilon_1$ can be found from the construction of Lyapunov function as given in Appendix I. 

Then for the system of equations \eqref{multimod02}, define a vector $\chi_1=[x_1,x_2]^T$ and a system of functions $\overline{F}_1(\chi_1)=[f_1(\cdot),f_2(\cdot)]^T$. A Lyapunov function $V_2(\chi_1,x_3)$ is chosen that satisfies the certain growth requirements for singularly perturbed system as in Appendix I as
\begin{equation}
\label{lyapn-1}
V_{2}(\chi_1,x_3)=(1-d_{2})V_{1}(\chi_1)+d_{2}W(\chi_1,x_3)
\end{equation}
where $0<d_{2}<1$ and $W(\chi_1,x_3)$ is the chosen Lyapunov function that satisfies the growth requirements for \eqref{multboundary}. 

\subsection{Three time scale behaviour of multiple groups of robots}
The matrix $\mathbf{P}$ and $\mathbf{R}$ of equation \eqref{eq25}, can be written in the following form
$$
\mathbf{P}=[\mathbf{P}_s^T, \mathbf{P}_r^T, \mathbf{P}_c^T]^T \ ; \
\mathbf{R}=[\mathbf{R}_s^T, \mathbf{R}_r^T, \mathbf{R}_c^T]^T
$$
Therefore, the collective dynamics of \eqref{eq25} can be separately written in the form of intra group shape dynamics $(Z_s)$, as follows
\begin{equation}
\label{eq101}
\ddot{Z}_s=\mathbf{P}_s \dot{\Lambda}+F_s+\mathbf{R}_s
\end{equation}
where, $
Z_s=\mathbf{\Psi_m} X; \ F_s=\mathbf{\Psi_m} BU
$. 
The inter group shape dynamics $(Z_r)$ is written as,
\begin{equation}
\label{eq102}
\ddot{Z}_r=\mathbf{P}_r \dot{\Lambda}+F_r+\mathbf{R}_r
\end{equation}
where, $
Z_r=\Psi_r X; \ F_r=\Psi_r BU
$. 
The dynamics of the overall leader $(z_c)$ is expressed as,
\begin{equation}
\label{eq103}
\ddot{z}_c=\mathbf{P}_c \dot{\lambda}+f_c+\mathbf{R}_c
\end{equation}
where, $z_c=\Psi_l X; \ f_l=\Psi_c BU$. \\
We define the intra group shape error vector as $Z_{se}=Z_s-Z_{sd}$, the inter group shape error vector $Z_{re}=Z_r-Z_{rd}$ , and the tracking error of overall leader $z_{ce}=z_c-z_{cd}$, where $Z_{sd}$, $Z_{rd}$, and $z_{cd}$ are the desired intra group, desired inter group shape variables, and desired trajectory of the overall leader respectively.\\
Define a set of three time instants $\tau_s, \tau_r$, and $\tau_c$, such that $\tau_s=\frac{\tau}{\epsilon_1\epsilon_2}$; $\tau_r=\frac{\tau}{\epsilon_1}$ $\tau_s \leq \tau_r \leq \tau_l \leq \infty$. The controller is to be designed such that $Z_{se} \rightarrow 0$, during the interval $[\tau_0,\tau_s]$, $Z_{re} \rightarrow 0$, during the interval $[\tau_0,\tau_r]$, $z_{ce} \rightarrow 0$, during the interval $[\tau_0,\tau_l]$. Here, $Z_{se}$ is ultra fast variable, $Z_{re}$ is fast variable, and $z_{ce}$ is slow variable. To achieve the desired formation and tracking, the following controllers is proposed for \eqref{eq101}-\eqref{eq103}.
\begin{equation}
\label{eq105}
\begin{split}
F_s &=\nu_s-\mathbf{P}_s \dot{Z}-\mathbf{R}_s+\ddot{Z}_{sd}\\
F_r &=\nu_r-\mathbf{P}_r \dot{Z}-\mathbf{R}_r+\ddot{Z}_{rd}\\
f_l &=\nu_l-\mathbf{P}_l \dot{z}-\mathbf{R}_l+\ddot{z}_{cd}\\
\end{split}
\end{equation}
where,
$$
\nu_s =- K_{s1} Z_{se}-K_{s2} \dot{Z}_{se}-\overline{K}_{sr} \dot{Z}_{re}-\overline{K}_{sc} \dot{z}_{ce}+\ddot{Z}_{sd}
$$
$$
\nu_r =-K_{r1} Z_{re}-K_{r2} \dot{Z}_{re}-\overline{K}_{rs} \dot{Z}_{se}-\overline{K}_{rc}\dot{z}_{ce}+\ddot{z}_{rd}
$$
$$
\nu_c =- k_{c1} z_{ce}- k_{c2} \dot{z}_{ce}-\overline{K}_{cs} \dot{Z}_{se}-\overline{K}_{cr}\dot{Z}_{re}+\ddot{z}_{cd}
$$
where, $K_{s1}=\frac{K_{fs1}}{(\epsilon_1 \epsilon_2)^2}$, 
%$K_{fs1}=k_{fs1}I$, 
$K_{s2}=\frac{K_{fs2}}{\epsilon_1 \epsilon_2}$, 
%$K_{fs2}=k_{fs2}I$, 
$K_{r1}=\frac{K_{fr1}}{\epsilon_1 ^2}$, 
%$K_{fr1}=k_{fr1}I$, 
$K_{r2}=\frac{K_{fr2}}{\epsilon_1}$, 
are controller gain matrices.
$K_{fs1}=k_{s1}I_{2N_s}$, $K_{fs2}=k_{s2}I_{2N_s}$ and $K_{r1}=k_{r1}I_{2N_r}$, $K_{r2}=k_{r2}I_{2N_r}$, where $N_s$ is the number of intra group shape variables, $N_r$ is the number of inter group shape variables, and $k_{s1},k_{s2}\in \mathbb{R}^+$, $k_{r1},k_{r2}\in \mathbb{R}^+$.
$\overline{K}$s are coupling gain matrices, where $\overline{K}_{sr}\in \mathbb{R}^{2N_s\times 2N_r}$, $\overline{K}_{sc}\in \mathbb{R}^{2N_s\times 2}$, $\overline{K}_{rs}\in \mathbb{R}^{2N_r\times 2N_s}$, $\overline{K}_{rc}\in \mathbb{R}^{2N_r\times 2}$, $\overline{K}_{cs}\in \mathbb{R}^{2\times 2N_r}$, $\overline{K}_{cr}\in \mathbb{R}^{2\times 2N_r}$.\\
Using \eqref{eq105}, the closed loop error dynamics is given as
\begin{equation}
\label{eq45}
\begin{split}
\ddot{Z}_{se} &=-K_{s1} Z_{se}-K_{s2} \dot{Z}_{se}-\overline{K}_{sr} \dot{Z}_{re}-\overline{K}_{sl} \dot{z}_{ce} \\
\ddot{Z}_{re} &=-K_{r1} Z_{re}-K_{r2} \dot{Z}_{re}-\overline{K}_{rs} \dot{\Lambda}_{se}-\overline{K}_{rc} \dot{z}_{ce} \\
\ddot{z}_{ce} &=-k_{l1} z_{ce}-k_{l2} \dot{z}_{ce}-\overline{K}_{cs} \dot{Z}_{se}-\overline{K}_{cr} \dot{Z}_{re}
\end{split}
\end{equation}
\textbf{Theorem 1} Suppose the controllers $F_s$, $F_r$, and $f_l$, as given in \eqref{eq105} is designed for system \eqref{eq45}. Then The system \eqref{eq45} is exponentially stable for all $\epsilon_i\leq\epsilon_i^*$, for some small $\epsilon_i^*$,  $i=1,2$. The analytical upper bounds on $\epsilon_i$, $i=1,2$ are derived to establish the stability of whole singularly perturbed system.\\
\textbf{Proof:} To write error dynamics define\\
$$
E_c=\begin{bmatrix}
z_{ce} \\
\dot{z}_{ce}
\end{bmatrix};
E_r=\begin{bmatrix}
\frac{1}{\epsilon_1}Z_{re} \\
\dot{Z}_{re}
\end{bmatrix};
E_s=\begin{bmatrix}
\frac{1}{\epsilon_1\epsilon_2}Z_{se} \\
\dot{Z}_{se}
\end{bmatrix}
$$
Hence, the error dynamics of \eqref{eq45} is written in the form of singularly perturbed system as follows
$$
\dot{E}_c=\begin{bmatrix}
0 & 1 \\
-k_{c1} & -k_{c2}
\end{bmatrix} E_c+\begin{bmatrix}
0 & 0 \\
0 & -\overline{K}_{cs} 
\end{bmatrix} E_s+
\begin{bmatrix}
0 & 0 \\
0 & -\overline{K}_{cr} 
\end{bmatrix} E_r
$$
$$
\epsilon_1 \dot{E}_r=\begin{bmatrix}
0 & I \\
-K_{fr1} & -K_{fr2}
\end{bmatrix} E_r+
\epsilon_1 \bigg ( \begin{bmatrix}
0 & 0 \\
0 & -\overline{K}_{rs} 
\end{bmatrix} E_s
$$
$$
+\begin{bmatrix}
0 & 0 \\
0 & -\overline{K}_{rc} 
\end{bmatrix} E_c \bigg )
$$
$$
\epsilon_1\epsilon_2\dot{E}_s=\begin{bmatrix}
0 & I \\
-K_{fs1} & -K_{fs2}
\end{bmatrix} E_s+
\epsilon_1\epsilon_2 \bigg ( 
\begin{bmatrix}
0 & 0 \\
0 & -\overline{K}_{sr} 
\end{bmatrix} E_r
$$
$$+\begin{bmatrix}
0 & 0 \\
0 & -\overline{K}_{sc} 
\end{bmatrix} E_c \bigg )
$$
Sequential application of two time scale results in bottom up approach sets $\epsilon_2=\epsilon_{1}=0$. As a result the slow manifolds become $E_s=0$, $E_r=0$. The boundary layer systems are derived as follows
\begin{equation}
\label{bdry1}
\frac{dE_s}{d\tau_i}= A_sE_s \ ; \ \tau_i=\frac{t}{\epsilon_1\epsilon_2}
\end{equation}
\begin{equation}
\label{bdry2}
\frac{dE_r}{d\tau_r}=A_r E_r \ ; \ \tau_r=\frac{t}{\epsilon_1}
\end{equation}
where, $A_s=\begin{bmatrix}
0 & I \\
-K_{fs1} & -K_{fs2}
\end{bmatrix}$ and $A_r=\begin{bmatrix}
0 & I \\
-K_{fr1} & -K_{fr2}
\end{bmatrix}$. As the above boundary layer systems are all linear and time invariant, the exponential stability can be guaranteed if matrices $A_s$ and $A_r$ are stable. Notice that these matrices are in companion form. So, there always exist a pair of gain matrices $(K_{fr1}, K_{fr2})$ and $(K_{fs1}, K_{fs2})$ in order to ensure the stability of $A_s$ and $A_r$.\\
We choose Lyapunov functions for the boundary layer \eqref{bdry1} as
\begin{equation}
\label{liapi}
V_s(E_s)=E_s^TP_sE_s
\end{equation}
where $Q_s>0$ such that matrices $P_s$ satisfies Lyapunov equations $A_s^TP_s+P_s^TA_s=-Q_s$ such that $\dot{V_s}(E_s)\leq 0$. Similarly for the boundary layer \eqref{bdry2}, the Lyapunov function is chosen as
\begin{equation}
\label{liapr}
V_r(E_r)=E_r^TP_rE_r
\end{equation}
where $Q_r>0$ such that matrix $P_r$ satisfies Lyapunov equation $A_r^TP_r+P_r^TA_r=-Q_r$ such that $\dot{V_r}(E_r)\leq 0$.
With proper choice of the gains $k_{c1}=k_1I_2$ and $k_{c2}=k_2I_2$, the reduced order slow system
\begin{equation}
\label{reduced}
\dot{E}_c=A_c E_c
\end{equation}
where, $A_c=\begin{bmatrix}
0 & 1 \\
-k_{c1} & -k_{c2}
\end{bmatrix}$ and the Lyapunov function for the reduced order system can be chosen as
\begin{equation}
\label{liapc}
V_c(E_c)=E_c^TP_cE_c
\end{equation}
Hence, the overall system is locally exponentially stable for small values of $\epsilon_1$ and $\epsilon_2$.\\
To derive the bounds on $\epsilon_i$, a composite Lyapunov function of the following form
\begin{equation}
\label{lyapcr}
V(E_c, E_r)=(1-d_1)V_c(E_l)+d_1V_r(E_r)
\end{equation}
where $0<d_1<1$, is chosen to satisfy the following condition 
\begin{equation}
\label{liap01}
\begin{split}
&\dot{V}(E_c,E_r)=-[(1-d_1)E^T_cQ_cE_c-d_1E^T_cA_{rc}^TP_rE_r \\
&-d_1E^T_rP_rA_{rc}E_r+\frac{d_1}{\epsilon_1}E_r^TQ_rE_r] \\
&=-\begin{bmatrix}
E_c \\
E_r
\end{bmatrix}^T \begin{bmatrix}
(1-d_1)Q_c & -d_1P_rA_{rc} \\
-d_1A_{rc}^TP_r & \frac{d_1}{\epsilon_1}Q_r
\end{bmatrix} \begin{bmatrix}
E_c \\
E_r
\end{bmatrix} \\
&\leq -\chi^T Q_{\epsilon_1} \chi
\end{split}
\end{equation}
where $\chi=[E_c^T,E_r^T]^T$, $A_{rc}=\begin{bmatrix}
0 & 0 \\
0 & -\overline{K}_{rc}
\end{bmatrix}$ and $Q_{\epsilon_1}>0$.
From $Q_{\epsilon_1}$, the bound on $\epsilon_1$ can be found using Schur's compliment for positive definiteness: $A>0$ and $C-B^TA^{-1}B>0$, where, $A=(1-d_1)Q_c$, $B=-d_1P_rA_{rc}$, and $C=\frac{d_1}{\epsilon_1}Q_r$ of the matrix $Q_{\epsilon_1}$. The explicit bound on $\epsilon_1$ is
\begin{equation}
\label{be1}
\epsilon_1<det(A)det(\overline{C}-B^TA^{-1}B)
\end{equation}
where $\overline{C}=\epsilon_1C$.

We then construct another Lyapunov function of the following form to find the composite stability of subsystems $\chi$ and $E_s$
\begin{equation}
\label{lyapcr}
V(\chi, E_s)=(1-d_2)V(\chi)+d_2V_s(E_s)
\end{equation}
The time derivative of \eqref{lyapcr} gives the following additional terms
\begin{equation}
\label{dlyapcr}
\begin{split}
\dot{V}(\chi, E_s)&=(1-d_2)\dot{V}(\chi)-\frac{d_2}{\epsilon_1\epsilon_2}E_s^TQ_sE_s \\
&+E_r^TA_{sr}P_sE_s+E_c^TA_{sc}^TP_sE_s \\
&+E_s^TP_sA_{sr}E_r+E_s^TP_sA_{sc}E_c \\
&\leq -\chi_1^T Q_{\epsilon_2} \chi_1
\end{split}
\end{equation}
which leads to the construction of another matrix $Q_{\epsilon_2}$ in the same way $Q_{\epsilon_1}$ is derived in \eqref{liap01}, and $\chi_1=[E_c^T,E_r^T,E_s^T]^T$, $A_{sr}=\begin{bmatrix}
0 & 0 \\
0 & -\overline{K}_{sr}
\end{bmatrix}$, and $A_{sc}=\begin{bmatrix}
0 & 0 \\
0 & -\overline{K}_{sc}
\end{bmatrix}$. The matrix $Q_{\epsilon_2}$ has the following form
\begin{equation}
\label{liap02}
\begin{bmatrix}
(1-d_1)(1-d_2)Q_l & d_1(1-d_2)P_rA_{rc} & -d_2A_{sc}^TP_s \\
-d_1(1-d_2)A_{rc}^TP_r & \frac{d_1}{\epsilon_1}(1-d_2)Q_r & -d_2A_{sr}^TP_s \\
-d_2P_sA_{sc} & -d_2P_sA_{sr} & \frac{d_2}{\epsilon_1\epsilon_2}Q_s
\end{bmatrix}
\end{equation}
Then the composite system is asymptotically stable if $Q_{\epsilon_2}>0$. The bound on $\epsilon_2$ can be computed from $Q_{\epsilon_2}$ using Schur's compliment for positive definiteness stated above. The matrices are $A=(1-d_2)Q_{\epsilon_1}$, $B=[-d_2P_sA_{sc}, -d_2P_sA_{sr}]^T$ and $C=\frac{d_2}{\epsilon_1\epsilon_2}Q_s$ of the matrix $Q_{\epsilon_2}$.
This completes the proof of \textit{theorem 1}.$\blacksquare$

\subsection{Asymptotic Stability Analysis of Multi-time Scale Singularly Perturbed Systems}

The asymptotic stability analysis of multi-time scale singularly perturbed systems composed of the repeated application of two-time scale analysis, given in Appendices I and II. There are two ways to address the analysis: \textit{Top-Down} and \textit{Bottom-Up}. These two approaches logically select the slow and fast dynamics sequentially. The details of these two approaches can be found in \cite{c32}-\cite{c44}. The \textit{Bottom-Up} approach is considered here in this paper to understand the natural evolution of the multi-time scale singularly perturbed systems in their own configuration spaces. A generic multi-time scale model can be described by
\begin{equation}
\label{multimod01}
\begin{split}
\dot{x}_1&=f_1(x_1,x_2,\cdots,x_n), x_1\in R^{m_1} \\
\epsilon_1\dot{x}_2&=f_2(x_1,x_2,\cdots,x_n), x_2\in R^{m_2} \\
&\vdots \\
(\prod_{i=1}^{n-1}\epsilon_{i})\dot{x}_n&=f_n(x_1,x_2,\cdots,x_n), x_n\in R^{m_n}
\end{split}
\end{equation}
The model \eqref{multimod01} can be sequentially decomposed into $(n-1)$ different two-time scale models. The first two-time scale model considers the time scale defined by the stretched time scale $t_{n-1}=\frac{t}{\prod_{i=1}^{n-1}\epsilon_{i}}$, where the reduced (slow) subsystem is defined by
\begin{equation}
\label{multimod02}
\begin{split}
\dot{x}_1&=f_1(x_1,x_2,\cdots,F_n(x_1,\cdots,x_{n-1})) \\
\epsilon_1\dot{x}_2&=f_2(x_1,x_2,\cdots,x_{n-1},F_n(x_1,\cdots,x_{n-1})) \\
&\vdots \\
(\prod_{i=1}^{n-2}\epsilon_{i})\dot{x}_{n-1}&=f_{n-1}(x_1,x_2,\cdots,x_{n-1},F_n(x_1,\cdots,x_{n-1}))
\end{split}
\end{equation}
and the boundary layer subsystem is given by
\begin{equation}
\label{multboundary}
\frac{dx_n}{dt_{n-1}}=f_n(x_1,x_2,\cdots,x_n(t_{n-1}))
\end{equation}
where in \eqref{multboundary}, $x_1,x_2,\cdots,x_{n-1}$ are treated like fix parameters and $x_n(t_{n-1})$ evolves on its stretched time scale $t_{n-1}$. $F_n(x_1,\cdots,x_{n-1})$ in \eqref{multimod02} represents the quasi-steady-state of the boundary layer \eqref{multboundary}, when $\epsilon_{n-1}=0$, that is, $0=f_n(x_1,x_2,\cdots,x_n)\rightarrow x_n=F_n(x_1,\cdots,x_{n-1})$. Then for the system of equations \eqref{multimod02}, define a vector $\chi_1=[x_1,\cdots,x_{n-1}]^T$ and a system of functions $\overline{F}_1(\chi_1,x_n)=[f_1(\cdot),\cdots, f_{n-1}(\cdot)]^T$. A Lyapunov function $V_{n-1}(\chi_1,x_n)$ is chosen that satisfies the growth requirements given in Appendix I and II as
\begin{equation}
\label{lyapn-1}
V_{n-1}(\chi_1,x_n)=(1-d_{n-1})V_{n-2}(\chi_1)+d_{n-1}W(\chi_1,x_n)
\end{equation}
where $0<d_{n-1}<1$ and $W(\chi_1,x_n)$ is the chosen Lyapunov function that satisfies the growth requirements for \eqref{multboundary}. Applying the same procedure leads to the construction of $V_{n-2}(\chi_1)=V_{n-2}(\chi_2,x_{n-1})$, $\chi_2=[x_1,\cdots,x_{n-2}]^T$. Note that the construction of $V_{n-2}(\chi_2,x_{n-1})$ with the Lyapunov functions for the subsystems $\chi_2$ and $x_{n-1}$ must satisfy the growth requirements on $\overline{F}_2(\chi_2,x_{n-1},F_n(x_1,\cdots,x_{n-1}))$ and $f_{n-1}(\chi_2,x_{n-1},F_n(x_1,\cdots,x_{n-1}))$. The bound on $\epsilon_n$ can be found from the construction of Lyapunov function as given in Appendix II. Similarly, the bounds on other $\epsilon_i$s can be found following the same procedure sequentially until we reach the reduced order slow system
\begin{equation}
\label{slowest}
\begin{split}
\dot{x}_1&=f_1(x_1,\cdots,F_{n-1}(x_1,\cdots,x_{n-2}, \\
&F_n(x_1,\cdots,x_{n-1})),F_n(x_1,\cdots,x_{n-1}))
\end{split}
\end{equation}

\subsection{Multi time scale behaviour of multiple groups of robots}

The objective of this section is to show that multi time scale convergence of the collective dynamics of \eqref{eq23} can be achieved in singular perturbation framework, depending upon the selection of gain parameters of the designed controller. \\
The matrix $\mathbf{P}$ and $\mathbf{R}$ of equation \eqref{eq25} in subsection A, can be written as $\mathbf{P}=[\mathbf{P}_1^T$, $\mathbf{P}_2^T$, $\hdots$, $\mathbf{P}_m^T$, $\mathbf{P}_r^T$, $\mathbf{P}_c^T$, $]^T$, and $\mathbf{R}=[\mathbf{R}_1^T$, $\mathbf{R}_2^T$, $\hdots$, $\mathbf{R}_m^T$, $\mathbf{R}_r^T$, $\mathbf{R}_c^T$, $]^T$.
Therefore, the collective dynamics of \eqref{eq23} can be separately written in the form of intra group shape dynamics $(Z_i)$, as follows
\begin{equation}
\label{eq203}
\ddot{Z}_i=\mathbf{P}_i \dot{Z}+F_i+\mathbf{R}_i
\end{equation}
where, $
Z_i=\Phi_i X; \ F_m=\Phi_i BU$, $i=1,2,..,m$. 
The inter group shape dynamics $(Z_r)$ is written as,
\begin{equation}
\label{eq204}
\ddot{Z}_r=\mathbf{P}_r \dot{Z}+F_r+\mathbf{R}_r
\end{equation}
where, $
Z_r=\Phi_r X; \ F_r=\Phi_r BU
$. 
The dynamics of the centroid $(z_c)$ is expressed as,
\begin{equation}
\label{eq205}
\ddot{z}_c=\mathbf{P}_c \dot{Z}+f_c+\mathbf{R}_c
\end{equation}
where, $z_c=\Phi_c X; \ f_c=\Phi_c BU$. \\
The intra group shape error vectors are defined by $Z_{1e}=Z_1-Z_{1d}, Z_{2e}=Z_2-Z_{2d},...,Z_{me}=Z_m-Z_{md}$ , where $Z_{1d}, Z_{2d},...,Z_{md}$ are the desired intra group shape vectors. The inter group shape error vectors are defined as $Z_{re}=Z_r-Z_{rd}$, and the tracking error of centroid is defined as $z_{ce}=z_c-z_{cd}$, where, $Z_{rd}$ and $z_{cd}$ are the desired inter group shape variables and desired trajectory of the centroid respectively.\\
Define a set of $(m+2)$ time instants $t_1,t_2,\ldots,t_m,t_r$, and $t_c$ such that
$t_1=\frac{t}{\epsilon_1\epsilon_2\cdots\epsilon_{m+1}}$; $t_2=\frac{t}{\epsilon_1\epsilon_2\cdots\epsilon_m}$;$\ldots$ $t_m=\frac{t}{\epsilon_1\epsilon_2}$; $t_r=\frac{t}{\epsilon_1}$; $t_1\leq t_2 \leq ... \leq t_m \leq t_ r \leq t_c \leq t$ as $t\rightarrow \infty$. $t$ is total time of operation, $t_1 \cdots t_r$ are stretched time scales (within which the subsystems must converge), and $\epsilon_1 \cdots \epsilon_{m+1}$ are controller gain parameters chosen to achieve different time scale convergence. The controllers are to be designed such that intra group shape error vectors $Z_{ie} \rightarrow 0$, during the interval $[t_0,t_i]$, $i=1, \ldots, m$. The inter group shape error $Z_{re}\rightarrow 0$, during the interval $[t_0,t_r]$, and tracking error $z_{ce} \rightarrow 0$, during the interval $[t_0,t_c]$. 
To achieve the desired formation the following controllers is proposed for \eqref{eq203} - \eqref{eq205}.
\begin{equation}
\label{eq206}
\begin{split}
F_i &=\nu_i-\mathbf{P}_i \dot{Z}-\mathbf{R}_i+\ddot{Z}_{id}, \ i=1, 2, \ldots, m.\\
F_r &=\nu_r-\mathbf{P}_r \dot{Z}-\mathbf{R}_r+\ddot{Z}_{rd}\\
f_c &=\nu_c-\mathbf{P}_c \dot{Z}-\mathbf{R}_c+\ddot{Z}_{cd}\\
\end{split}
\end{equation}
where,
\begin{equation}
\label{eqnus}
\begin{split}
\nu_i &=-K_{i1} Z_{ie}-K_{i2} \dot{Z}_{ie}-\sum_{j=1,j\neq i}^{m} \overline{K}_{ij} \dot{Z}_{je}-\overline{K}_{ir} \dot{Z}_{re} \\ 
&-\overline{K}_{ic} \dot{Z}_{ce} + \ddot{Z}_{id}, \ i=1, 2, \ldots, m. \\
\nu_r &=-K_{r1} Z_{re}-K_{r2} \dot{Z}_{re}-\sum_{j=1}^{m} \overline{K}_{rj} \dot{Z}_{je} -\overline{K}_{rc} \dot{Z}_{ce} + \ddot{Z}_{rd} \\
\nu_c &=-k_{c1} Z_{ce}-k_{c2} \dot{Z}_{ce}-\sum_{j=1}^{m} \overline{K}_{cj} \dot{Z}_{je}-\overline{K}_{cr} \dot{Z}_{re}+\ddot{z}_{cd}
\end{split}
\end{equation}
where, $K_{i1}=\frac{K_{fi1}}{(\prod_{i=1}^m\epsilon_{i+1})^2}$, $K_{i2}=\frac{K_{fi2}}{(\prod_{i=1}^m\epsilon_{i+1})}$, $i=1,..,,m+1$, 
%$K_{m1}=\frac{K_{fm1}}{(\epsilon_1\epsilon_2)^2}$, $K_{m2}=\frac{K_{fm2}}{\epsilon_1\epsilon_2}$, 
$K_{r1}=\frac{K_{fr1}}{\epsilon_1^2}$, $K_{r2}=\frac{K_{fr2}}{\epsilon_1}$ are controller gain matrices. Suppose, there are $N_s$ intra group shape variables and $N_r$ inter group shape variables for $m$ groups of robots. Define a set $N_s=\{N_{s1},N_{s2},...,N_{sm}\}$ to denote the number of intra group shape variables in each group. 
%Then 
%\begin{equation}
%\label{gainval}
%\begin{split}
%&K_{fi1}=k_{i1}I_{2N_{si}} ;\ K_{fi2}=k_{i2}I_{2N_{si}},i=1,...,m \\
%&K_{r1}=k_{r1}I_{2N_r} ;\ K_{r1}=k_{r2}I_{2N_r} \\
%&k_{c1}=k_1I_2 ;\ k_{c2}=k_2I_2
%\end{split}
%\end{equation}
%where, $k_{i1},k_{i2},i=1,...,m,k_{r1},k_{r2},k_1,k_2\in \mathbb{R}^+$.
$\overline{K}$s are coupling gain matrices, where $\overline{K}_{ij} \in \mathbb{R}^{2N_{si}\times 2N_{sj}}$, $i,j=1,2,...,m$ and $i \neq j$; $\overline{K}_{ir}\in \mathbb{R}^{2N_{si}\times 2N_r}$, $i=1,2,...,m$; $\overline{K}_{ic}\in \mathbb{R}^{2N_{si}\times 2}$, $i=1,2,...,m$; $\overline{K}_{rj}\in \mathbb{R}^{2N_r\times 2N_{sj}}$, $j=1,2,...,m$; $\overline{K}_{rc}\in \mathbb{R}^{2N_r\times 2}$, $\overline{K}_{cj}\in \mathbb{R}^{2\times 2N_{sj}}$, $j=1,2,...,m$;  $\overline{K}_{cr}\in \mathbb{R}^{2\times 2N_r}$.
Hence, applying the control law into \eqref{eq25}, the closed loop error dynamics are given as follows
\begin{equation}
\label{eq207}
\begin{split}
\ddot{Z}_{ie} &=-K_{i1} Z_{ie}-K_{i2} \dot{Z}_{ie}-\sum_{j=1,j\neq i}^{m} \overline{K}_{ij} \dot{Z}_{je}-\overline{K}_{ir} \dot{Z}_{re} \\ 
& \ \ \ -\overline{K}_{ic} \dot{Z}_{ce}, \ i=1,2, \ldots, m\\
\ddot{Z}_{re} &=-K_{r1} Z_{re}-K_{r2} \dot{Z}_{re}-\sum_{j=1}^{m} \overline{K}_{rj} \dot{Z}_{je} -\overline{K}_{rc} \dot{Z}_{ce}\\
\ddot{z}_{ce} &=-k_{c1} Z_{ce}-k_{c2} \dot{Z}_{ce}-\sum_{j=1}^{m} \overline{K}_{cj} \dot{Z}_{je}-\overline{K}_{cr} \dot{Z}_{re}
\end{split}
\end{equation}
The main results of this paper for the convergence of error dynamics \eqref{eq207} is stated in the form of the following theorem. As a result, the intra group shape variables $Z_i$, $i=1,...,m$, inter group shape variables $Z_r$ converge to their desired values in different time scales. Also the centroid $z_c$ converges to the desired trajectory.\\
\textbf{Theorem 2} Suppose the controllers $F_1$, $\ldots$, $F_m$, $F_r$, and $f_c$, as given in \eqref{eq206} is designed for system \eqref{eq207}. Then The system \eqref{eq207} is exponentially stable for all $\epsilon_i\leq\epsilon_i^*$, for some small $\epsilon_i^*$,  $i=1,2,...,m+1$. Under mild conditions stated in Appendix A-E, the analytical upper bounds on $\epsilon_i$, $i=1,2,...,m+1$ are derived to establish the stability of whole singularly perturbed system.\\
%(ii) Let inequalities of Appendix A-E be satisfied. Then there exists analytical upper bounds on $\epsilon_i$, $i=1,2,...,m$.\\
%\textbf{Theorem 1} The controllers $F_1$, $\ldots$, $F_m$, $F_r$, and $f_c$, as given in \eqref{eq206} locally exponentially stabilize system \eqref{eq207}, for all $\epsilon_1 < \epsilon_1^*$, $\epsilon_2 < \epsilon_2^*$, $\ldots$, $\epsilon_{m+1} < \epsilon_{m+1}^*$ and for some small $\epsilon_1^*$, $\epsilon_2^*$, $\ldots$, $\epsilon_{m+1}^*$. Moreover, the closed loop system is asymptotically decoupled. As a result, the intra group shape variable $Z_1$, $\ldots$, $Z_m$ inter group shape variables $Z_r$ converge to their desired values in different time scale. Also the centroid $z_c$ converges to the desired trajectory.\\
%\textbf{Theorem 1} Let inequalities of Appendix A-E be satisfied. Then the origin is an asymptotically stable equilibrium of the singularly perturbed system () for all $\epsilon_i\in(0,\epsilon_1^*)$. Moreover,  for every number $d_i\in(0,1)$, the resulting Lyapunov function is a Lyapunov function for all $\epsilon_1(0,\epsilon_{d_1})$, where $\epsilon_i\leq\epsilon_i^*$.
\textbf{Proof:} The proof consists of two parts. In the first part the stability of the reduced and boundary layer systems, is analyzed by a Lyapunov function, which is given by the composition of the Lyapunov functions of the slow and the fast systems. The analytical bounds on singularly perturbed parameters $\epsilon_i$ are derived in the second part.\\
To write error dynamics define
$$
E_c=\begin{bmatrix}
Z_{ce} \\
\dot{Z}_{ce}
\end{bmatrix};
E_r=\begin{bmatrix}
\frac{1}{\epsilon_1}Z_{re} \\
\dot{Z}_{re}
\end{bmatrix};
E_i=\begin{bmatrix}
\frac{1}{\prod_{i=1}^m\epsilon_{i+1}}Z_{ie} \\
\dot{Z}_{ie}
\end{bmatrix} 
$$
Hence, the error dynamics of the equation \eqref{eq207} is written in the form of singularly perturbed system as follows
\begin{equation}
\begin{split}
\dot{E}_c&=\begin{bmatrix}
0 & I \\
-k_{c1} & -k_{c2}
\end{bmatrix} E_c+\sum_{i=1}^m \begin{bmatrix}
0 & 0 \\
0 & -\overline{K}_{ci} 
\end{bmatrix} E_i \\
&+\begin{bmatrix}
0 & 0 \\
0 & -\overline{K}_{cr} 
\end{bmatrix} E_r
\end{split}
\end{equation}
\begin{equation}
\begin{split}
\epsilon_1 \dot{E}_r&=\begin{bmatrix}
0 & I \\
-K_{fr1} & -K_{fr2}
\end{bmatrix} E_r+\epsilon_1 \bigg ( \sum_{i=1}^m\begin{bmatrix}
0 & 0 \\
0 & -\overline{K}_{ri} 
\end{bmatrix} E_i \\
&+\begin{bmatrix}
0 & 0 \\
0 & -\overline{K}_{rc} 
\end{bmatrix} E_c \bigg )
\end{split}
\end{equation}
\begin{equation}
\begin{split}
(\prod_{i=1}^m\epsilon_{i+1})\dot{E}_i&=\begin{bmatrix}
0 & I \\
-K_{fi1} & -K_{fi2}
\end{bmatrix} E_i+
(\prod_{i=1}^m\epsilon_{i+1})\\
&\bigg ( \sum_{j=1,k \neq i}^m\begin{bmatrix}
0 & 0 \\
0 & -\overline{K}_{ij} 
\end{bmatrix} E_j+ 
\begin{bmatrix}
0 & 0 \\
0 & -\overline{K}_{ir} 
\end{bmatrix} E_r \\
&+\begin{bmatrix}
0 & 0 \\
0 & -\overline{K}_{ic}
\end{bmatrix} E_c \bigg )
\end{split}
\end{equation}
Sequential application of two time scale results in bottom up approach sets $\epsilon_{m+1}=\epsilon_{m}=...=\epsilon_{1}=0$. As a result the slow manifolds become $E_m=0$,$\cdots$, $E_1=0$, $E_r=0$. The boundary layer systems are derived as follows
\begin{equation}
\label{bdry1}
\frac{dE_i}{dt_i}= A_iE_i \ ; \ t_i=\frac{t}{\prod_{i=1}^{m}\epsilon_{i+1}}
\end{equation}
\begin{equation}
\label{bdry2}
\frac{dE_r}{dt_r}=A_r E_r \ ; \ t_r=\frac{t}{\epsilon_1}
\end{equation}
where, $A_i=\begin{bmatrix}
0 & I \\
-K_{fi1} & -K_{fi2}
\end{bmatrix}$, $i=1,2,...,m$ and $A_r=\begin{bmatrix}
0 & I \\
-K_{fr1} & -K_{fr2}
\end{bmatrix}$. As the above boundary layer systems are all linear and time invariant, the exponential stability can be guaranteed if matrices $A_i$ and $A_r$ are stable. Notice that these matrices are in companion form. So, there always exist a pair of gain matrices $(K_{fr1}, K_{fr2})$, $(K_{fm1}, K_{fm2})$, $\cdots$, $(K_{f11},K_{f12})$ in order to ensure the stability of $A_i$ and $A_r$.\\
We choose Lyapunov functions for the boundary layer \eqref{bdry1} as
\begin{equation}
\label{liapi}
V_i(E_i)=E_i^TP_iE_i
\end{equation}
where $Q_i>0$ such that matrices $P_i$ satisfies Lyapunov equations $A_i^TP_i+P_i^TA_i=-Q_i$ for $i=1,2,...,m$. Similarly for the boundary layer \eqref{bdry2}, the Lyapunov function is chosen as
\begin{equation}
\label{liapr}
V_r(E_r)=E_r^TP_rE_r
\end{equation}
where $Q_r>0$ such that matrix $P_r$ satisfies Lyapunov equation $A_r^TP_r+P_r^TA_r=-Q_r$.
With proper choice of the gains $k_{c1}=k_1I_2$ and $k_{c2}=k_2I_2$, the reduced order slow system
\begin{equation}
\label{reduced}
\dot{E}_c=A_c E_c
\end{equation}
where, $A_c=\begin{bmatrix}
0 & 1 \\
-k_{c1} & -k_{c2}
\end{bmatrix}$ and the Lyapunov function for the reduced order system can be chosen as
\begin{equation}
\label{liapc}
V_c(E_c)=E_c^TP_cE_c
\end{equation}
Hence, the overall system is asymptotically stable for small values of $\epsilon_1, \epsilon_2, \cdots, \epsilon_{m+1}$.\\
\textbf{Remark 1} 
The computation of the respective bounds on $\epsilon_i$, $i=1,2$ is clarified here for the three time scale dynamics of \cite{c39}. The same procedure is to be extended to find further bounds on $\epsilon_i$, $i=3,...,m$ for multi-time scale dynamics. To derive the bounds on $\epsilon_1$, a composite Lyapunov function of the following form
\begin{equation}
\label{lyapcr}
V(E_c, E_r)=(1-d_1)V_c(E_l)+d_1V_r(E_r)
\end{equation}
where $0<d_1<1$, is chosen to satisfy the following condition 
\begin{equation}
\label{liap01}
\begin{split}
&\dot{V}(E_c,E_r)=-[(1-d_1)E^T_cQ_cE_c-d_1E^T_cA_{rc}^TP_rE_r \\
&-d_1E^T_rP_rA_{rc}E_r+\frac{d_1}{\epsilon_1}E_r^TQ_rE_r] \\
&=-\begin{bmatrix}
E_c \\
E_r
\end{bmatrix}^T \begin{bmatrix}
(1-d_1)Q_c & -d_1P_rA_{rc} \\
-d_1A_{rc}^TP_r & \frac{d_1}{\epsilon_1}Q_r
\end{bmatrix} \begin{bmatrix}
E_c \\
E_r
\end{bmatrix} \\
&\leq -\chi^T Q_{\epsilon_1} \chi
\end{split}
\end{equation}
where $\chi=[E_c^T,E_r^T]^T$, $A_{rc}=\begin{bmatrix}
0 & 0 \\
0 & -\overline{K}_{rc}
\end{bmatrix}$ and $Q_{\epsilon_1}>0$.
From $Q_{\epsilon_1}$, the bound on $\epsilon_1$ can be found using Schur's compliment for positive definiteness: $A>0$ and $C-B^TA^{-1}B>0$, where, $A=(1-d_1)Q_c$, $B=-d_1P_rA_{rc}$, and $C=\frac{d_1}{\epsilon_1}Q_r$ of the matrix $Q_{\epsilon_1}$. The explicit bound on $\epsilon_1$ is
\begin{equation}
\label{be1}
\epsilon_1<det(A)det(\overline{C}-B^TA^{-1}B)
\end{equation}
where $\overline{C}=\epsilon_1C$.

We then construct another Lyapunov function of the following form to find the composite stability of subsystems $\chi$ and $E_s$
\begin{equation}
\label{lyapcr}
V(\chi, E_s)=(1-d_2)V(\chi)+d_2V_s(E_s)
\end{equation}
The time derivative of \eqref{lyapcr} gives the following additional terms
\begin{equation}
\label{dlyapcr}
\begin{split}
\dot{V}(\chi, E_s)&=(1-d_2)\dot{V}(\chi)-\frac{d_2}{\epsilon_1\epsilon_2}E_s^TQ_sE_s \\
&+E_r^TA_{sr}P_sE_s+E_c^TA_{sc}^TP_sE_s \\
&+E_s^TP_sA_{sr}E_r+E_s^TP_sA_{sc}E_c \\
&\leq -\chi_1^T Q_{\epsilon_2} \chi_1
\end{split}
\end{equation}
which leads to the construction of another matrix $Q_{\epsilon_2}$ in the same way $Q_{\epsilon_1}$ is derived in \eqref{liap01}, and $\chi_1=[E_c^T,E_r^T,E_s^T]^T$, $A_{sr}=\begin{bmatrix}
0 & 0 \\
0 & -\overline{K}_{sr}
\end{bmatrix}$, and $A_{sc}=\begin{bmatrix}
0 & 0 \\
0 & -\overline{K}_{sc}
\end{bmatrix}$. The matrix $Q_{\epsilon_2}$ has the following form
\begin{equation}
\label{liap02}
\begin{bmatrix}
(1-d_1)(1-d_2)Q_l & d_1(1-d_2)P_rA_{rc} & -d_2A_{sc}^TP_s \\
-d_1(1-d_2)A_{rc}^TP_r & \frac{d_1}{\epsilon_1}(1-d_2)Q_r & -d_2A_{sr}^TP_s \\
-d_2P_sA_{sc} & -d_2P_sA_{sr} & \frac{d_2}{\epsilon_1\epsilon_2}Q_s
\end{bmatrix}
\end{equation}
Then the composite system is asymptotically stable if $Q_{\epsilon_2}>0$. The bound on $\epsilon_2$ can be computed from $Q_{\epsilon_2}$ using Schur's compliment for positive definiteness stated above. The matrices are $A=(1-d_2)Q_{\epsilon_1}$, $B=[-d_2P_sA_{sc}, -d_2P_sA_{sr}]^T$ and $C=\frac{d_2}{\epsilon_1\epsilon_2}Q_s$ of the matrix $Q_{\epsilon_2}$.

\section{Collision Avoidance}

The controllers of \eqref{eq206} don't guarantee collision avoidance among the robots. Therefore, the barrier-like function of \cite{c19} is chosen as a potential function for collision avoidance. The modified form of the function for the robots $i,j\in \mathbb{N}$, $i,j=1,2,...,N$ is given by
\begin{equation}
\label{potf}
V_{ij}(p_i,p_j)=\bigg( min\left\{0,\frac{\parallel q_i-q_j\parallel^2 - R^2}{\parallel q_i-q_j\parallel^2 - r^2}\right\}\bigg )^2
\end{equation}
where $R$ is the radius of sensing and $q_i,q_j$ represents the position of $i$-th and $j$th robot respectively. $r$ denotes the permissible distance from the robot $i$ to avoid collision. Then the control input for the collision avoidance of $i$-th robot is the summation of all potential defined by \eqref{potf} of the robots $j$ inside the permissible distance $r$:
\begin{equation}
\label{pforce}
\bigtriangledown f_i=-\sum_{j=1,j\neq i}^{n} \frac{\partial V_{ij}(p_i,p_j)}{\partial p_i} ^T
\end{equation}
where $\frac{\partial y}{\partial x}$ is the gradient of a scalar function $y$ (of dependent ($x$) and independent variables) with respect to $x$ and $\bigtriangledown f_i\in\mathbb{R}^{2\times 1}$. Define a matrix $\bigtriangledown F\in\mathbb{R}^{2N\times 1}$ of control input based on avoidance potential of all robots $i=1,2,...,N$ as
\begin{equation}
\label{totf}
\bigtriangledown F=[\bigtriangledown f_1^T,\bigtriangledown f_2^T,...,\bigtriangledown f_N^T]^T
\end{equation}
To comply with the solutions of $\ddot{X}=-\bigtriangledown F$ under the transformation $Z=\Phi_M X$, define a vector of control input in the transformed domain as
\begin{equation}
\label{tp}
F_{pot}=\Phi_M \bigtriangledown F
\end{equation}

\subsection{Three Time Scale}

The vector $F_{pot}$ of \eqref{tp} is partitioned as $F_{pot}=[F_{pot\bold{s}}^T,F_{pot\bold{r}}^T,F_{pot\bold{c}}^T]^T$, where, $F_{pot\bold{s}}\in \mathbb{R}^{2\rho \times 1}$, $F_{pot\bold{r}}\in \mathbb{R}^{2(m-1)\times 1}$, and $F_{pot\bold{c}}\in \mathbb{R}^{2\times 1}$. Then the equations for formation controllers with collision avoidance is given by 
\begin{equation}
\label{pbc3t}
\begin{split}
F_s &=\nu_s-\mathbf{P}_s \dot{Z}-\mathbf{R}_s+\ddot{Z}_{sd}-k_sF_{pot\bold{s}}\\
F_r &=\nu_r-\mathbf{P}_r \dot{Z}-\mathbf{R}_r+\ddot{Z}_{rd}-k_rF_{pot\bold{r}}\\
f_c &=\nu_c-\mathbf{P}_c \dot{Z}-\mathbf{R}_c+\ddot{Z}_{cd}-k_cF_{pot\bold{c}}\\
\end{split}
\end{equation}
$k_s=\frac{1}{\epsilon_1\epsilon_2},k_r=\frac{1}{\epsilon_1},k_c=1$ are scalars associated with the potential terms $F_{pot\bold{s}},F_{pot\bold{r}},F_{pot\bold{c}}$ respectively to adjust the gain. Hence, the closed loop dynamics are
\begin{equation}
\label{pbccl3t}
\begin{split}
\ddot{Z}_{se} &=-K_{s1} Z_{se}-K_{s2} \dot{Z}_{se}-\overline{K}_{sr} \dot{Z}_{re}-\overline{K}_{sc} \dot{z}_{ce}-k_sF_{pot\bold{s}} \\
\ddot{Z}_{re} &=-K_{r1} Z_{re}-K_{r2} \dot{Z}_{re}-\overline{K}_{rs} \dot{Z}_{se}-\overline{K}_{rc} \dot{z}_{ce}-k_rF_{pot\bold{r}} \\
\ddot{z}_{ce} &=-k_{c1} z_{ce}-k_{c2} \dot{z}_{ce}-\overline{K}_{cs} \dot{Z}_{se}-\overline{K}_{cr} \dot{Z}_{re}-k_cF_{pot\bold{c}}
\end{split}
\end{equation}
\textbf{Theorem 3} The controllers $F_s$, $F_r$, and $f_c$, as given in \eqref{eq105} locally asymptotically stabilize system \eqref{eq45}, independently, for all $\epsilon_i < \epsilon_i^*$, $i=1,2$ and for some small $\epsilon_i^*$, $i=1,2$.
%, i.e., the time of convergence will not depend upon $\epsilon_i^*$, $i=1,2$. 
As a result, the intra group shape variable $Z_s$, inter group shape variables $Z_r$, and the centroid $z_c$ converge to their desired values as $t\rightarrow\infty$.\\
\textbf{Proof:} To write error dynamics define
$$
E_c=\begin{bmatrix}
z_{ce} \\
\dot{z}_{ce}
\end{bmatrix};
E_r=\begin{bmatrix}
\frac{1}{\epsilon_1}Z_{re} \\
\dot{Z}_{re}
\end{bmatrix};
E_s=\begin{bmatrix}
\frac{1}{\epsilon_1\epsilon_2}Z_{se} \\
\dot{Z}_{se}
\end{bmatrix}
$$

Hence, the error dynamics of \eqref{eq45} is written in the form of singularly perturbed system as follows
$$
\dot{E}_c=\begin{bmatrix}
0 & 1 \\
-k_{c1} & -k_{c2}
\end{bmatrix} E_c+\begin{bmatrix}
0 & 0 \\
0 & -\overline{K}_{cs} 
\end{bmatrix} E_s
$$
$$
+
\begin{bmatrix}
0 & 0 \\
0 & -\overline{K}_{cr} 
\end{bmatrix} E_r-
\begin{bmatrix}
0 \\
F_{pot\bold{c}} 
\end{bmatrix} 
$$
$$
\epsilon_1 \dot{E}_r=\begin{bmatrix}
0 & I \\
-K_{fr1} & -K_{fr2}
\end{bmatrix} E_r+
\epsilon_1 \bigg ( \begin{bmatrix}
0 & 0 \\
0 & -\overline{K}_{rs} 
\end{bmatrix} E_s
$$
$$
+\begin{bmatrix}
0 & 0 \\
0 & -\overline{K}_{rc} 
\end{bmatrix} E_c \bigg )-
\begin{bmatrix}
0 \\
F_{pot\bold{r}} 
\end{bmatrix} 
$$
$$
\epsilon_1\epsilon_2\dot{E}_s=\begin{bmatrix}
0 & I \\
-K_{fs1} & -K_{fs2}
\end{bmatrix} E_s+
\epsilon_1\epsilon_2 \bigg ( 
\begin{bmatrix}
0 & 0 \\
0 & -\overline{K}_{sr} 
\end{bmatrix} E_r
$$
$$
+\begin{bmatrix}
0 & 0 \\
0 & -\overline{K}_{sc} 
\end{bmatrix} E_c \bigg )-
\begin{bmatrix}
0 \\
F_{pot\bold{s}} 
\end{bmatrix} 
$$
By setting $\epsilon_1=\epsilon_2=0$, we've the slow manifolds $E_r=\begin{bmatrix}
0 & I \\
-K_{fr1} & -K_{fr2}
\end{bmatrix}^{-1}\begin{bmatrix}
0 \\
F_{pot\bold{r}} 
\end{bmatrix}$, $E_s=\begin{bmatrix}
0 & I \\
-K_{fs1} & -K_{fs2}
\end{bmatrix}^{-1}\begin{bmatrix}
0 \\
F_{pot\bold{s}} 
\end{bmatrix} $. And the boundary layer systems are as follows
$$
\frac{dE_r}{dt_r}=\begin{bmatrix}
0 & I \\
-K_{fr1} & -K_{fr2}
\end{bmatrix} E_r-\begin{bmatrix}
0 \\
F_{pot\bold{r}} 
\end{bmatrix} \ ; \ t_r=\frac{t}{\epsilon_1}
$$
$$
\frac{dE_s}{dt_s}=\begin{bmatrix}
0 & I \\
-K_{fs1} & -K_{fs2}
\end{bmatrix} E_s-\begin{bmatrix}
0 \\
F_{pot\bold{s}} 
\end{bmatrix} \ ; \ t_s=\frac{t}{\epsilon_1 \epsilon_2}
$$
As the above boundary layer systems are a combination of a dissipative part and a linear combination of potentially decreasing functions, the subsystems will exponentially reach the trajectory of the gradient of the potential term within the time $\tau_1$ and $\tau_2$ respectively. So if there is a possibility of collision among the robots after $\tau_1$ and $\tau_2$, the formation will collapse as the potential terms aren't time dependent functions. Due to the effect of the potential terms in the boundary layers, the subsystems again reach desired formation when the inter robot distance criteria are met.\\
With proper choice of the gains $k_{c1}$ and $k_{c2}$, for example, $k_{c1}=k_{c2}=kI_2$ where $k$ is a scalar, the reduced order slow system
$$
\dot{E}_c=\begin{bmatrix}
0 & 1 \\
-k_{c1} & -k_{c2}
\end{bmatrix} E_c
$$
$$+\begin{bmatrix}
0 & 0 \\
0 & -\overline{K}_{cs} 
\end{bmatrix}\begin{bmatrix}
0 & I \\
-K_{fs1} & -K_{fs2}
\end{bmatrix}^{-1}\begin{bmatrix}
0 \\
F_{pot\bold{s}} 
\end{bmatrix}
$$
$$+\begin{bmatrix}
0 & 0 \\
0 & -\overline{K}_{cr} 
\end{bmatrix}\begin{bmatrix}
0 & I \\
-K_{fr1} & -K_{fr2}
\end{bmatrix}^{-1}\begin{bmatrix}
0 \\
F_{pot\bold{r}} 
\end{bmatrix}+\begin{bmatrix}
0 \\
F_{pot\bold{c}} 
\end{bmatrix}
$$
is also exponentially stable because the gains multiplied with the potential term only add to the total potential. Thus the potential terms preserve the property of driving away the neighbouring robots to avoid collision. Hence, the overall system is stable for small values of $\epsilon_1, \epsilon_2$.$\blacksquare$\\
\textbf{Remark} It's necessary for the robots not to collide at the time of intra group formation or inter group formation or tracking the given trajectory. Hence, the potential term is required for the fast system when they reach the boundary layer, because it's important to avoid collision even when the subsystems reach the desired formation. For example, after the convergence of intra groups formation, there is a fair possibility that the formed groups collide at time time of inter group formation. Hence the potential terms are added with adjustable scalar gains to the controllers in \eqref{pbc3t}.

\subsection{Multi Time Scale}
The vector $F_{pot}$ of \eqref{tp} is partitioned as $F_{pot}=[F_{pot\bold{1}}^T,F_{pot\bold{2}}^T,...,F_{pot\bold{m}}^T,F_{pot\bold{r}}^T,F_{pot\bold{c}}^T]^T$, where, $F_{pot\bold{s}}\in \mathbb{R}^{2\rho \times 1}$, $F_{pot\bold{r}}\in \mathbb{R}^{2(m-1)\times 1}$, and $F_{pot\bold{c}}\in \mathbb{R}^{2\times 1}$. Then \eqref{eq206} is modified as below to assure collision avoidance
\begin{equation}
\label{pbcmt}
\begin{split}
F_i &=\nu_i-\mathbf{P}_i \dot{Z}-\mathbf{R}_i+\ddot{Z}_{id}-k_iF_{pot\textbf{i}}, \ i=1, 2, \ldots, m.\\
F_r &=\nu_r-\mathbf{P}_r \dot{Z}-\mathbf{R}_r+\ddot{Z}_{rd}-k_rF_{pot\textbf{r}}\\
f_c &=\nu_c-\mathbf{P}_c \dot{Z}-\mathbf{R}_c+\ddot{Z}_{cd}-k_cF_{pot\textbf{c}}\\
\end{split}
\end{equation}
The scalars $k_1=\frac{1}{\epsilon_1\epsilon_2},k_2=\frac{1}{\epsilon_1\epsilon_2\epsilon_3},...,k_m=\frac{1}{\epsilon_1\epsilon_2...\epsilon_{m+1}},k_r=\frac{1}{\epsilon_1},k_c=1$ are associated with the potential terms $F_{pot\bold{1}},F_{pot\bold{2}},...,F_{pot\bold{m}},F_{pot\bold{r}},F_{pot\bold{c}}$ respectively to adjust the performance of the controllers. Hence, the closed loop dynamics are
\begin{equation}
\label{pbcclmt}
\begin{split}
\ddot{Z}_{ie} &=-K_{i1} Z_{ie}-K_{i2} \dot{Z}_{ie}-\sum_{j=1,j\neq i}^{m} \overline{K}_{ij} \dot{Z}_{je}-\overline{K}_{ir} \dot{Z}_{re} \\ 
&-\overline{K}_{ic} \dot{Z}_{ce}-k_iF_{pot\textbf{i}}, \ i=1,2, \ldots, m\\
\ddot{Z}_{re} &=-K_{r1} Z_{re}-K_{r2} \dot{Z}_{re}-\sum_{j=1}^{m} \overline{K}_{rj} \dot{Z}_{je} -\overline{K}_{rc} \dot{Z}_{ce}\\
&-k_rF_{pot\textbf{r}}\\
\ddot{z}_{ce} &=-k_{c1} Z_{ce}-k_{c2} \dot{Z}_{ce}-\sum_{j=1}^{m} \overline{K}_{cj} \dot{Z}_{je}-\overline{K}_{cr} \dot{Z}_{re}\\
&-k_cF_{pot\textbf{c}}
\end{split}
\end{equation}
\textbf{Theorem 4} The controllers $F_1$, $\ldots$, $F_m$, $F_r$, and $f_c$, as given in \eqref{pbcmt} asymptotically stabilize system \eqref{pbcclmt}, for all $\epsilon_1 < \epsilon_1^*$, $\epsilon_2 < \epsilon_2^*$, $\ldots$, $\epsilon_{m+1} < \epsilon_{m+1}^*$ and for some small $\epsilon_1^*$, $\epsilon_2^*$, $\ldots$, $\epsilon_{m+1}^*$. Moreover, the closed loop system is asymptotically decoupled. As a result, the intra group shape variable $Z_1$, $\ldots$, $Z_m$ inter group shape variables $Z_r$ converge to their desired values in different time scale. Also the centroid $z_c$ converges to the desired trajectory.\\
\textbf{Proof:} To write error dynamics define
$$
E_c=\begin{bmatrix}
Z_{ce} \\
\dot{Z}_{ce}
\end{bmatrix};
E_r=\begin{bmatrix}
\frac{1}{\epsilon_1}Z_{re} \\
\dot{Z}_{re}
\end{bmatrix};
E_i=\begin{bmatrix}
\frac{1}{\prod_{i=1}^m\epsilon_{i+1}}Z_{ie} \\
\dot{Z}_{ie}
\end{bmatrix} 
$$
Hence, the error dynamics of the equation \eqref{eq207} is written in the form of singularly perturbed system as follows
\begin{equation}
\begin{split}
\dot{E}_c&=\begin{bmatrix}
0 & 1 \\
-k_{c1} & -k_{c2}
\end{bmatrix} E_c+\sum_{i=1}^m \begin{bmatrix}
0 & 0 \\
0 & -\overline{K}_{ci} 
\end{bmatrix} E_i \\
&+\begin{bmatrix}
0 & 0 \\
0 & -\overline{K}_{cr} 
\end{bmatrix} E_r-\begin{bmatrix}
0 \\
F_{pot\textbf{c}}
\end{bmatrix}
\end{split}
\end{equation}
\begin{equation}
\begin{split}
\epsilon_1 \dot{E}_r&=\begin{bmatrix}
0 & 1 \\
-K_{fr1} & -K_{fr2}
\end{bmatrix} E_r+\epsilon_1 \bigg ( \sum_{i=1}^m\begin{bmatrix}
0 & 0 \\
0 & -\overline{K}_{ri} 
\end{bmatrix} E_i \\
&+\begin{bmatrix}
0 & 0 \\
0 & -\overline{K}_{rc} 
\end{bmatrix} E_c \bigg )-\begin{bmatrix}
0 \\
F_{pot\textbf{r}}
\end{bmatrix}
\end{split}
\end{equation}
\begin{equation}
\begin{split}
(\prod_{i=1}^m\epsilon_{i+1})\dot{E}_i=\begin{bmatrix}
0 & 1 \\
-K_{fi1} & -K_{fi2}
\end{bmatrix} E_i+
(\prod_{i=1}^m\epsilon_{i+1}) \\
\bigg ( \sum_{j=1,k \neq i}^m\begin{bmatrix}
0 & 0 \\
0 & -\overline{K}_{ij} 
\end{bmatrix} E_j+ 
\begin{bmatrix}
0 & 0 \\
0 & -\overline{K}_{ir} 
\end{bmatrix} E_r \\
+\begin{bmatrix}
0 & 0 \\
0 & -\overline{K}_{ic}
\end{bmatrix} E_c \bigg )-\begin{bmatrix}
0 \\
F_{pot\textbf{m}}
\end{bmatrix}
\end{split}
\end{equation}
For each equation condition 1 and condition 3 of \textit{Theorem 4} is satisfied. By setting $\epsilon_1=\epsilon_2=...=\epsilon_{m+1}=0$, we've the boundary layer equations as follows
$$
\frac{dE_r}{dt_r}=\begin{bmatrix}
0 & 1 \\
-K_{fr1} & -K_{fr2}
\end{bmatrix} E_r-\begin{bmatrix}
0 \\
F_{pot\textbf{r}}
\end{bmatrix} \ ; \ t_r=\frac{t}{\epsilon_1}
$$
$$
\frac{dE_i}{dt_i}=\begin{bmatrix}
0 & 1 \\
-K_{fi1} & -K_{fi2}
\end{bmatrix} E_i-\begin{bmatrix}
0 \\
F_{pot\textbf{i}}
\end{bmatrix}; \ t_i=\frac{t}{\prod_{i=1}^{m}\epsilon_{i+1}}
$$
As the above boundary layer systems are a combination of a dissipative part and a linear combination of potentially decreasing functions, the subsystems will exponentially reach the trajectory of the gradient of the potential term within the time $\tau_1$ and $\tau_2$ respectively. So if there is a possibility of collision among the robots after $\tau_1$ and $\tau_2$, the formation will collapse as the potential terms aren't time dependent functions. Due to the effect of the potential terms in the boundary layers, the subsystems again reach desired formation when the inter robot distance criteria are met.\\
With proper choice of the gains $k_{c1}=I_2$ and $k_{c2}=I_2$, the reduced order slow system
$$
\dot{E}_c=\begin{bmatrix}
0 & 1 \\
-k_{c1} & -k_{c2}
\end{bmatrix} E_c
$$
$$+\sum_{i=1}^{m} \begin{bmatrix}
0 & 0 \\
0 & -\overline{K}_{ci} 
\end{bmatrix}\begin{bmatrix}
0 & I \\
-K_{fi1} & -K_{fi2}
\end{bmatrix}^{-1}\begin{bmatrix}
0 \\
F_{pot\bold{i}} 
\end{bmatrix}
$$
$$+\begin{bmatrix}
0 & 0 \\
0 & -\overline{K}_{cr} 
\end{bmatrix}\begin{bmatrix}
0 & I \\
-K_{fr1} & -K_{fr2}
\end{bmatrix}^{-1}\begin{bmatrix}
0 \\
F_{pot\bold{r}} 
\end{bmatrix}+\begin{bmatrix}
0 \\
F_{pot\bold{c}} 
\end{bmatrix}
$$
is also asymptotically stable because the gains multiplied with the potential term only add to the total potential. Thus the potential terms preserve the property of driving away the neighboring robots to avoid collision. Hence, the overall system is stable for small values of $\epsilon_1, \epsilon_2,...,\epsilon_{m+1}$.$\blacksquare$\\
\textbf{Remark 2} It's necessary for the robots not to collide at the time of intra group formation or inter group formation or tracking the given trajectory. Hence, the potential term is required for the fast system when they reach the boundary layer, because it's important to avoid collision even when the subsystems reach the desired formation. For example, after the convergence of intra groups formation, there is a fair possibility that the formed groups collide at time time of inter group formation. Hence the potential terms are added with adjustable scalar gains to the controllers in \eqref{pbcclmt}.

\section{SIMULATION RESULTS}

The controllers developed in section V-VI have been simulated on three groups of robots with three robots in each group. Three time scale convergence being an example of multi time scale convergence, is demonstrated in this section with simulation. The controller gain parameters are chosen as $K_{fr1}=K_{fr2}=kI_4$, $K_{f11}=K_{f12}=kI_{12}$, where, $k=1$, and $\epsilon_1=0.1$, $\epsilon_2=0.1$. The matrices $\overline{K}_{sr}$, $\overline{K}_{sc}$, $\overline{K}_{rs}$, $\overline{K}_{rc}$, $\overline{K}_{cs}$, $\overline{K}_{cr}$ are chosen to be all $1$s with appropriate dimension, so that the system becomes tightly coupled, although the degree of coupling is left as a choice for the user. 
All the figures in this section show the trajectories of the robots moving in formation. The positions of the robots are marked by '$\triangleright$' and each group contains three robots marked with red, green and blue color. Potential force parameters are taken from \cite{c19}. 
The desired trajectory of the centroid of the formation is kept as $z_c=[t;30sin(0.1t)]$. 
The rest of the desired vectors are framed as illustrated in Section IV. 
The desired shape of individual group is an equilateral triangle and also desired shape of the bigger triangle that tangles all the groups, is equilateral triangle. From Fig. 1, each side of the small triangle and the big triangle are $b=7m$ and $a=20m$ respectively. 
The shape variables in the transformed domain, are given as 
$Z_{sd}=[(-4.9497, 0)$, $(0, 6.0622)$, $(-4.9497, 0)$, $(0, 6.0622)$, $(-4.9497, 0)$, $(0, 6.0622)]^T$, $Z_{rd}=[(-14.1421, 0)$, $(0, 17.3205)]^T$.
\begin{figure}[H]
\centering
\fbox{\includegraphics[width=8cm, height=5cm]{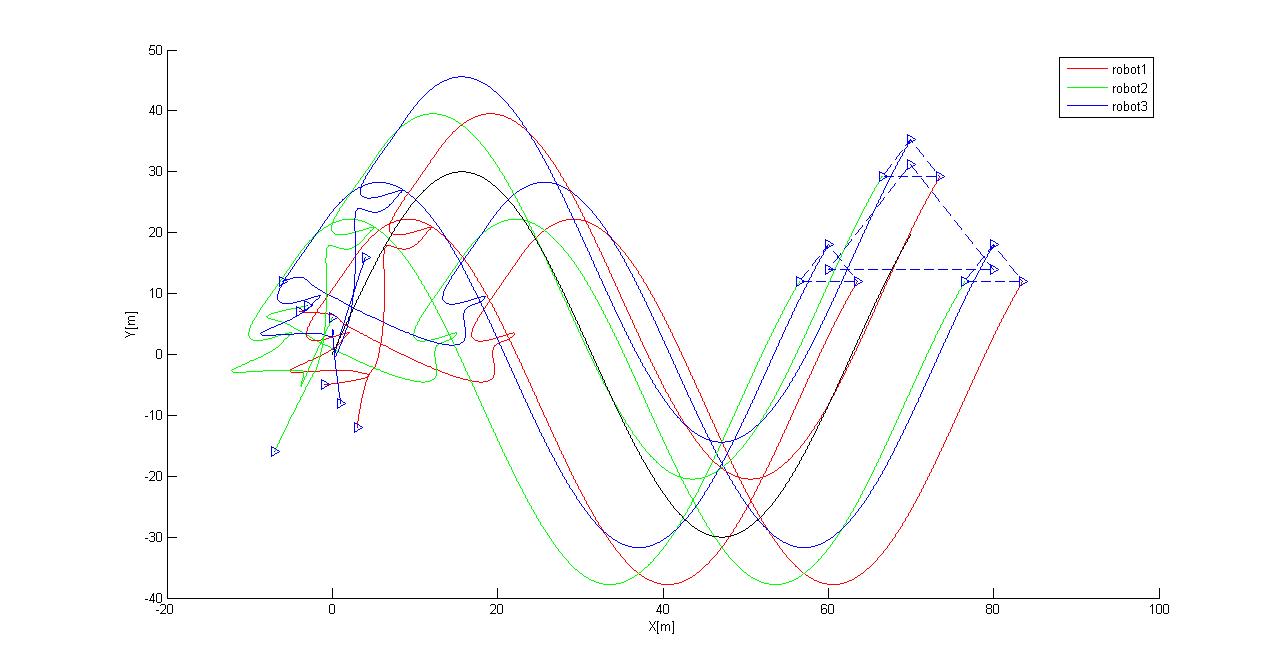}}
\caption{Formation control using transformation $\Phi_{M}$}
\end{figure}
\begin{figure}[H]
\centering
\fbox{\includegraphics[width=8cm, height=5cm]{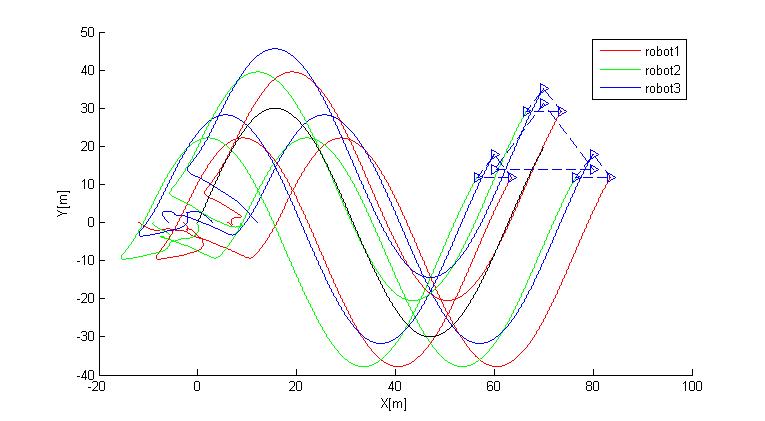}}
\caption{Potential force based Formation control using transformation $\Phi_{M}$}
\end{figure}

The initial values of position of $9$ robots is respectively as follows
$(x_1,y_1)=(-4,7)$, $(x_2,y_2)=,(-3,8)$, $(x_3,y_3)=(-6,12)$, $(x_4,y_4)=(-1,-5)$, $(x_5,y_5)=(0,6)$, $(x_6,y_6)=(1,-8)$, $(x_7,y_7)=(3,-12)$, $(x_8,y_8)=(-7,-16)$, $(x_9,y_9)=(4,16)$. In Fig. 2, it is shown that the robots converge to the desired formation from the initial conditions given above. Potential force has not been considered for the simulation in Fig. 2. The convergence of robots to the desired formation with collision avoidance, is depicted in Fig. 3.\\
Fig. 4 shows the convergence time of the states in the transformed domain separately (without applying potential force). All the intra group shape variables $Z_1 \ldots Z_6$ converge faster than inter group shape variables $Z_7$ and $Z_8$. It can also be seen from the figures, that the convergence of the centroid is the slowest of all. It is evident from Fig. 4 that the intra group shape variables converge to desired value at $t=0.1sec$. The inter group shape variables converge at time $t=1sec$ and the trajectory of centroid converges to the desired value at $t=10sec$. Thus convergence of intra group shape variables are $10$ times faster than the convergence of inter group shape variables. Again, convergence of the trajectory of centroid is $10$ times faster than the convergence of inter group shape variables. 
\begin{figure}[H]
\centering
\fbox{\includegraphics[width=8cm, height=6cm]{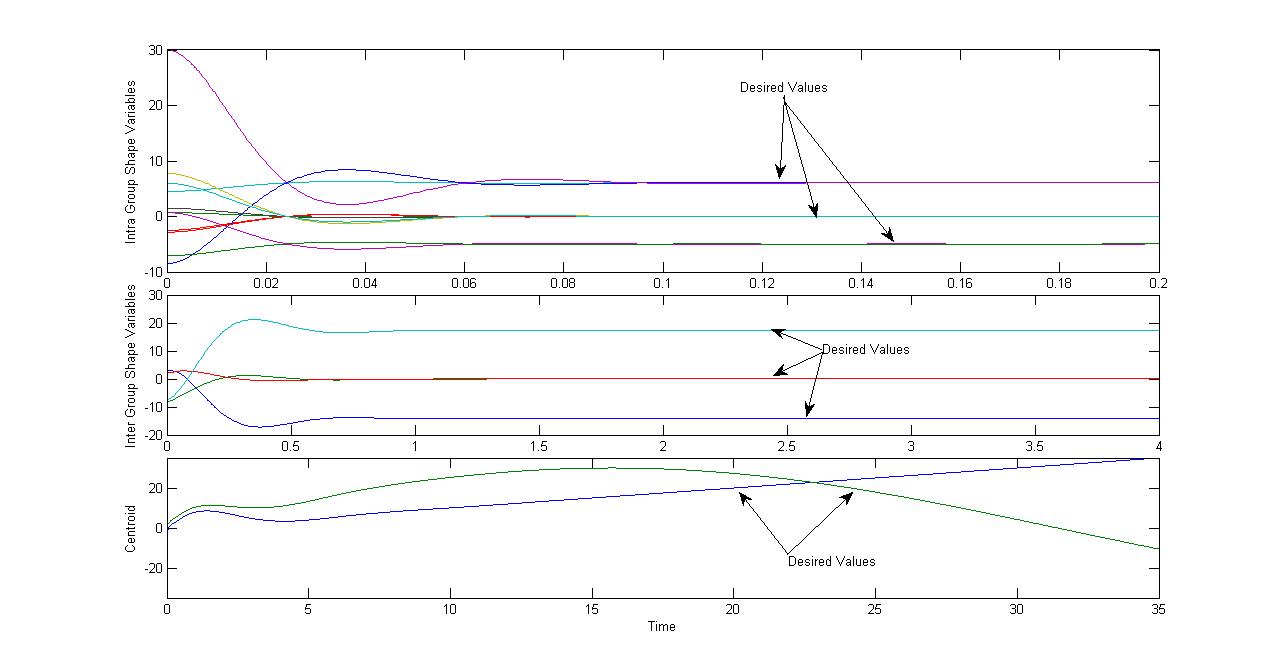}}
\caption{Plot of intra and inter group shape variables and centroid vs time}
\end{figure}

\section{CONCLUSION}
In the paper we propose, an intuitive and simple way of solving a complicated formation control problem. For that a centroid based transformation is given for multiple groups of robots such that a modular architecture results, in the form of intra group, inter group shape variables, and centroid. Thus exploiting the modularity separate controllers have been designed for each module of formation. The gains of the feedback controllers are so selected that the error dynamics become singularly perturbed system and multi time scale behaviour of the overall system is achieved. Thus the control laws ensure different time scale convergence of different group of robot dynamics. For collision avoidance, negative gradient of potential function has also been appended the proposed feedback controller. Simulation results shows the performance of the proposed formation controllers. It is to be noted that the high gains are to be chosen such that the input energy does not exceed the maximum capacity of the motor attached to the wheels of the WMR. A new class of problems for solving the group formation of multiple groups of robots, has been originated due to this formulation, and many potential formation problems can be solved using the proposed methodology. Future work entails multiple group formation of quadrotor robots with in the singularly perturbed system framework.

\appendices
\section{Asymptotic Stability Analysis Of Two-Time Scale Singular Perturbation Autonomous Systems}
To preserve the integrity of the paper, this section describes the general formulation for the asymptotic stability analysis for the two-time scale singular perturbation problems \cite{c41}, \cite{c42}. It is the foundation of the analysis that was conducted to demonstrate the asymptotic stability of the multi-time scale model in Section V. C. Consider a nonlinear autonomous singular perturbed system of the form
\begin{equation}
\label{spslow}
\dot{x}=f(x,z), \ x \in R^n,
\end{equation}
\begin{equation}
\label{spfast}
\epsilon\dot{z}=g(x,z), \ x \in R^m,
\end{equation}
which has an isolated equilibrium at the origin $(x=0, z=0)$. It is assumed throughout the formulation that $f$ and $g$ are smooth to ensure that for specified initial conditions, system \eqref{spslow} and \eqref{spfast} has a unique solution. The stability of the equilibrium is investigated by examining the reduced (slow) system
\begin{equation}
\label{spreduced}
\dot{x}=f(x,h(x))
\end{equation}
where $z=h(x)$ is an associated root of $0=g(x,z)$ and the boundary layer (fast) system
\begin{equation}
\label{spboundary}
\frac{dz}{d\tau}=g(x,z(\tau)), \ \tau=\frac{t}{\epsilon}
\end{equation}
where $x$ is treated as a fixed parameter and $\epsilon$ is the parasitic constant that defines the stretched time scale of the fast subsystem. To prove asymptotic stability, Lyapunov functions are to exist for reduced and boundary layer systems, which satisfy certain growth conditions to be addressed later. We assume that the following conditions hold for all
\begin{equation}
\label{cond1}
(x,y,\epsilon)\in [t_0,\infty)\times B_x \times B_z \times [0,\epsilon_1]
\end{equation}
where $\ B_x \in R^n$ and $B^z \in R^m$ denotes closed sets. We add and substract $f(x,h(x))$ to the right-hand side of \eqref{spslow} yielding
\begin{equation}
\label{sp01}
\dot{x}=f(x,h(x))+f(x,z)-f(x,h(x))
\end{equation}
where $f(x,z)-f(x,h(x))$ can be viewed as a perturbation of reduced system \eqref{spreduced}. It is natural first to look for a Lyapunov function candidate for \eqref{spreduced} and then to consider the effect of the perturbation term $f(x,z)-f(x,h(x))$ \cite{c41}.
\subsection{Assumption (Asymptotic Stability of the Origin)}
The origin $(x=0, z=0)$ is a unique and isolated equilibrium of \eqref{spslow} and \eqref{spfast}, that is,
\begin{equation}
\label{sporigin}
0=f(0,0) \ \text{and} \ 0=g(0,0);
\end{equation}
moreover, $z=h(x)$ is the unique root of $0=g(x,z,0)$ in $B^x \times B^z$, that is, $0=g(x,h(x))$, and there exists a class $\kappa$ function $p(\cdot)$ such that $\parallel h(x)\parallel\leq p(\parallel x \parallel)$.
To study the asymptotic stability of the equilibrium, Lyapunov function candidates are to be constructed for both reduced and boundary layer systems separately. The respective growth requirements will be defined separately in Assumptions B and C, whereas the growth requirements that combine both reduced and boundary layer system requirements, called interconnection conditions, will be defined in Assumptions D and E.
\subsection{Assumption (Reduced System Conditions)}
There exists a positive-definite Lyapunov function candidate $V(x)$, that is,
\begin{equation}
\label{ineqr01}
0<q_1(\parallel x\parallel)\leq V(x) \leq q_2(\parallel x\parallel)
\end{equation}
for some class $\kappa$ function $q_3(\cdot)$ and $q_4(\cdot)$, that satisfies the following inequality
\begin{equation}
\label{ineqr02}
\frac{\partial V}{\partial x}f(x,h(x))\leq -\alpha_1 \psi^2(x)
\end{equation}
where $\psi(\cdot)$ is a scalar function of vector arguments that vanishes only when its argument are zero and satisfying that $x=0$ is a stable equilibrium of the reduced order system. Condition \eqref{ineqr02} guarantees that $x=0$ is an asymptotically stable equilibrium of \eqref{spreduced}.
\subsection{Assumption (Boundary Layer System Conditions)}
There exists a positive-definite Lyapunov function candidate $W(x,z)$ such that for all $(x,z)\in B^x\times B^z$, satisfying
\begin{equation}
\label{ineq01}
0<q_3(\parallel z-h(x)\parallel)\leq W(x,z) \leq q_4(\parallel z-h(x)\parallel)
\end{equation}
for some class $\kappa$ function $q_3(\cdot)$ and $q_4(\cdot)$, that satisfies
\begin{equation}
\label{ineq02}
W(x,z)>0, \ \forall z \neq h(x)\ \text{and} \ W(x,h(x))=0
\end{equation}
and
\begin{equation}
\label{ineq03}
\frac{\partial W}{\partial z}g(x,z)\leq -\alpha_2 \phi^2(z-h(x)), \ \alpha_2>0
\end{equation}
where $W(x,z)$ is a Lyapunov function of boundary layer system \eqref{spboundary} in which $x$ is treated as a fixed parameter and $\phi(\cdot)$ is a scalar function of vector arguments that vanishes only when its argument are zero and satisfying that $z-h(x)$ is a stable equilibrium of the boundary layer system.
\subsection{Assumption (First Interconnection Condition)}
$V(x)$ and $W(x,z)$ must satisfy the so called interconnection conditions. The first interconnection condition is obtained by computing the derivative of $V$ along the solution of \eqref{sp01},
\begin{equation}
\label{lyap1}
\begin{split}
\dot{V}&=\frac{\partial V}{\partial x} f(x,h(x))+\frac{\partial V}{\partial x}[f(x,z)-f(x,h(x))] \\
&\leq \alpha_1 \psi^2(x)+\frac{\partial V}{\partial x}[f(x,z)-f(x,h(x))]
\end{split}
\end{equation}
assuming that
\begin{equation}
\label{ineq04}
\frac{\partial V}{\partial x}[f(x,z)-f(x,h(x))]\leq \beta_1\psi(x)\phi(z-h(x))
\end{equation}
so that
\begin{equation}
\label{lyapineq}
\dot{V}\leq -\alpha_1\phi^2(x)+\beta_1\psi(x)\phi(z-h(x))
\end{equation}
Inequality \eqref{ineq04} determines the allowed growth of $f$ in $z$, and $\alpha_1$ and $\beta_1$ are nonnegative constants.
\subsection{Assumption (Second Interconnection Conditions)}
The second interconnection condition is defined by
\begin{equation}
\label{ineqsec}
\frac{\partial W}{\partial x}f(x)\leq \gamma\phi^2(z-h(x))+\beta_2\psi(x)\phi(z-h(x)),
\end{equation}
where $\psi(\cdot)$ and $\phi(\cdot)$ are scalar functions of vector arguments that vanish only when their arguments are zero, that is, $\phi(x)=0$ if and only if $x=0$. $\gamma$ and $\beta_2$ are nonnegative constants.\\
With the Lyapunov function $V(x)$ and $W(x,z)$ obtained, a new Lyapunov function $\nu(x,z)$ is
considered and defined by the weighted sum of $V(x)$ and $W(x,z)$,
\begin{equation}
\label{lyapsing}
\nu(x,z)=(1-d)V(x)+dW(x,z)
\end{equation}
for $0<d<1$. $\nu(x,z)$ becomes the Lyapunov function candidate for the singular perturbed system \eqref{spslow}-\eqref{spfast}.
\subsection{Theorem A}
If $x=0$ is an asymptotically stable equilibrium of reduced system \eqref{spreduced}, $z=h(x)$ is an asymptotically stable equilibrium of boundary layer system \eqref{spboundary} uniformly in $x$, that is, the $\epsilon-\delta$ definition of Lyapunov stability and the convergence $z\rightarrow h(x)$ are uniform in $x$ \cite{c43}, and if $f(\cdot,\cdot)$ and $g(\cdot,\cdot)$ satisfy \eqref{ineqr02}, \eqref{ineq03}, \eqref{lyapineq}, and \eqref{ineqsec}, then the origin is an asymptotically stable equilibrium of the singularly perturbed system \eqref{spslow}, for sufficiently small $\epsilon$ \cite{c41}.
\section{Selection of The Bounds of the Stability Parameters}
Calculating the time derivative of $\nu$ of \eqref{lyapsing} along the trajectory of the full system \eqref{spslow}-\eqref{spfast}, we obtain
\begin{equation}
\label{lyap4proof}
\begin{split}
\dot{\nu}&=(1-d)\frac{\partial V}{\partial x}f(x,h(x))+\frac{d}{\epsilon}\frac{\partial W}{\partial z}g(x,z) \\
&+(1-d)\frac{\partial V}{\partial x}[f(x,z)-f(x,h(x))]+d\frac{\partial W}{\partial x}f(x)
\end{split}
\end{equation}
where using the \textit{Assumption B-E} of Appendix I, we can express \eqref{lyap4proof} as
\begin{equation}
\label{lyap4proof2}
\begin{split}
\dot{\nu}&\leq -(1-d)\alpha_1 \psi^2(x)+(1-d)\beta_1\psi(x)\phi(z-h(x))\\
&-\frac{d}{\epsilon}\alpha_2 \phi^2(z-h(x))+d\gamma\phi^2(z-h(x))\\
&+d\beta_2\psi(x)\phi(z-h(x))
=\begin{bmatrix}
\psi(x)\\
\phi(z-h(x))
\end{bmatrix}^T \times \\
&\begin{bmatrix}
(1-d)\alpha_1 & -\frac{1}{2}(1-d)\beta_1-\frac{1}{2}d\beta_2 \\
-\frac{1}{2}(1-d)\beta_1-\frac{1}{2}d\beta_2 & d(\frac{\alpha_2}{\epsilon}-\gamma)
\end{bmatrix} \\
&\times \begin{bmatrix}
\psi(x)\\
\phi(z-h(x))
\end{bmatrix}
\end{split}
\end{equation}
The right-hand side of (87) is a quadratic form in the comparison functions $\psi(x)$ and $\phi(z-h(x))$. The quadratic form is negative definite when
\begin{equation}
\label{condneg}
d(1-d)\alpha_1(\frac{\alpha_2}{\epsilon}-\gamma)>\frac{1}{4}[(1-d)\beta_1+d\beta_2]^2
\end{equation}
where $\alpha_1$, $\alpha_2$, $\beta_1$, $\beta_2$ and $\gamma$ are defined in \eqref{lyapineq} and \eqref{ineqsec}. Thus, rewriting \eqref{condneg} as
\begin{equation}
\label{condeps}
\epsilon<\frac{\alpha_1\alpha_2}{\alpha_1\gamma_1+\frac{1}{4(1-d)d}[(1-d)\beta_1+d\beta_2]^2}\equiv\epsilon_d
\end{equation}
Inequality \eqref{condeps} shows that for any choice of $d$, the corresponding $\nu$ is a Lyapunov function for the singular perturbed system \eqref{spslow}-\eqref{spfast} for all $\epsilon$ satisfying \eqref{condeps}. The maximum value of $\epsilon_d$ is given by
\begin{equation}
\label{epsval}
\epsilon^*=\frac{\alpha_1\alpha_2}{\alpha_1\gamma+\beta_1\beta_2}
\end{equation}
and occurs for
\begin{equation}
\label{dval}
d^*=\frac{\beta_1}{\beta_1+\beta_2}
\end{equation}
If all the growth requirements are satisfied, then the origin is an asymptotically stable equilibrium of the singularly perturbed system \eqref{spslow}-\eqref{spfast} for all $\epsilon\in(0,\epsilon^*)$, where $\epsilon^*$ is given by \eqref{dval}.

\end{document}